\documentclass[%
    reprint,superscriptaddress,
    nofootinbib,
    amsmath,amssymb,
    aps,pra,
    floatfix,
]{revtex4-2}

\usepackage{comment}
\usepackage{graphicx} % Include figure files
\usepackage{xcolor}
\usepackage{bm} % bold math
\usepackage{physics}
\usepackage{amssymb,amsmath,amsfonts,mathtools}
\usepackage{amsthm}
\usepackage{bbm}
\usepackage{hyperref}% add hypertext capabilities
\hypersetup{
    colorlinks,
    citecolor=blue,
    filecolor=blue,
    linkcolor=blue,
    urlcolor=blue
}

\newcommand{\sgn}{\operatorname{sgn}}
\newcommand{\pos}{\Theta}

\newtheorem*{proposition}{Proposition}

\makeatletter
\def\maketitle{
\@author@finish
\title@column\titleblock@produce
\suppressfloats[t]}
\makeatother

\begin{document}

\title{Tsirelson's Inequality for the Precession Protocol\texorpdfstring{\\}{ }is Maximally Violated by Quantum Theory}

\author{Lin Htoo Zaw}
\affiliation{Centre for Quantum Technologies, National University of Singapore, 3 Science Drive 2, Singapore 117543}

\author{Mirjam Weilenmann}
\affiliation{D\'epartement de Physique Appliqu\'ee, Universit\'e de Gen\`eve, Gen\`eve, Switzerland}

\author{Valerio Scarani}
\affiliation{Centre for Quantum Technologies, National University of Singapore, 3 Science Drive 2, Singapore 117543}
\affiliation{Department of Physics, National University of Singapore, 2 Science Drive 3, Singapore 117542}

\begin{abstract}
The precession protocol involves measuring $P_3$, the probability that a uniformly precessing observable (like the position of a harmonic oscillator or a coordinate undergoing spatial rotation) is positive at one of three equally spaced times. Tsirelson's inequality, which states that $P_3 \leq 2/3$ in classical theory, is violated in quantum theory by certain states. In this Letter, we address some open questions about the inequality: \emph{What is the maximum violation of Tsirelson's inequality possible in quantum theory? Might other theories do better?} By considering the precession protocol in a theory-independent manner for systems with finitely many outcomes, we derive a general bound for the maximum possible violation. This theory-independent bound must be satisfied by any theory whose expectation values are linear functions of observables---which includes classical, quantum, and all general probabilistic theories---and depends only on the minimum positive and negative measurement outcomes. Given any such two values, we prove by construction that quantum theory always saturates this bound. Some notable examples include the angular momentum of a spin-$3/2$ particle and a family of observables that outperform the quantum harmonic oscillator in the precession protocol. Finally, we also relate our findings to the recently introduced notion of constrained conditional probabilities.
\end{abstract}

\phantomsection\addcontentsline{toc}{part}{Tsirelson's Inequality for the Precession Protocol is Maximally Violated by Quantum Theory}
\maketitle

\section{Introduction}
Given that the dynamics of a harmonic oscillator is the same in both classical and quantum theory---a uniform precession in phase space---it is remarkable that one can \emph{positively} rule out classical theory using only position measurements at different times. The first such protocol was invented by Tsirelson in an overlooked preprint: he proved that $P_3$, the probability that a uniformly precessing observable is positive at one of three equally spaced times, satisfies the inequality $P_3 \leq 2/3$ in classical theory \cite{OG_tsirelson}. Tsirelson's inequality can be violated by some states of the quantum harmonic oscillator, thereby certifying their nonclassicality. Similar protocols that witness nonclassicality of spin angular momentum \cite{precession_protocol} and anharmonic dynamics \cite{precession_protocol_anharmonic}, and entanglement in composite systems~\cite{precession_protocol_CV_ent,precession_protocol_DV_ent}, have also been described recently.

The uniform precession stated above is an evolution of the type $\cos(\omega t) X + \sin(\omega t) Y$. In the case that $X$ and $Y$ are compatible observables, they could be simultaneously measured and thus admit a joint probability distribution. Their precession dynamics can therefore be simulated by a classical oscillator prepared with the same initial distribution, which cannot violate Tsirelson's inequality. \textit{A contrario}, the violation of Tsirelson's inequality certifies the impossibility of even assigning a joint probability distribution to the values of $X$ and $Y$, a stronger manifestation of incompatibility than that guaranteed by most uncertainty relations.

While quantum states with significant violation of Tsirelson's inequality have already been identified, it is natural to ask about the maximum violation achievable by quantum theory. \emph{Can quantum theory achieve the algebraic maximum $P_3 = 1$? Can any theory?} These questions parallel those in the study of quantum and post-quantum correlations with Bell inequalities, with the former answered by Tsirelson's bound \cite{Tsirelson-bound} and the latter by the Popescu-Rohrlich box \cite{PR_tsirelson,PR_paper}.

As for general theories, recent efforts in quantum foundations have aimed at identifying features of quantum theory that make it unique and powerful for applications in computing or information theory. Such ideas~\cite{Lee_2015, self-testing1, self-testing2} have been pursued in the framework of generalized probabilistic theories (GPTs)~\cite{Hardy2001, Barrett2007}, where notions of incompatibility have also been studied \cite{incompatibility-1,incompatibility-2,incompatibility-3}. In addition to aiding our understanding of quantum theory, these efforts may also lead to identifying new applications of quantum systems in the future, or provide tests that can rule out  postclassical theories different from quantum theory.

In this Letter, we give first evidence that quantum systems perform optimally under uniform precession, in the sense that they can maximally violate Tsirelson's inequality. Specifically, we derive a general theory-independent bound for the precession protocol, valid for all precessing discrete-variable systems. For this, we require that a theory has a notion of an observable that captures a measurement process, from which the possible measurement outcomes and outcome probabilities can be obtained from some description of the state of the system. Our general bound then applies under a few natural assumptions on the observables: that taking the mean value of a linear combination of observables is the same as the linear combination of the mean values; that the set of measurement outcomes is finite; and that the observables satisfy an algebraic equation that describes a uniform precession at the times of the measurements. All finite-dimensional GPTs satisfy these assumptions, as do additional theories. Then we show that the general bound is saturated by quantum theory: in particular, the previously studied case of the angular momentum of a spin-$3/2$ particle saturates this bound. Thus, we have identified a sense in which the incompatibility of observables allowed by quantum theory is maximal.

Finally, we provide further insights into the differences between precession of classical and quantum systems by analyzing the example of a clock, where the classical and quantum versions have the same spectrum, and by analyzing the probability spaces of the quantum observables studied in this work.

\section{Tsirelson's Original Inequality}
Let us present the original inequality by Tsirelson, already with a more general notation and relaxing some assumptions with respect to his work. Consider two observables $X(t)$ and $Y(t)$ that vary with time: these would be functions that output the values of $X$ or $Y$ at time $t$ for the classical case, and operators in the Heisenberg picture for the quantum case. We say that the pair $(X,Y)$ precesses uniformly over the times $\{t_k = kT/3\}_{k\in\mathbbm{Z}}$ with period $T$ if they satisfy
\begin{equation}\label{eq:precession-condition}
\begin{aligned}
    X_k \coloneqq X(t_k) &= \cos(\frac{2\pi k}{3}) X(t_0) + \sin(\frac{2\pi k}{3}) Y(t_0),\\
    Y_k \coloneqq  Y(t_k) &= \cos(\frac{2\pi k}{3}) Y(t_0) - \sin(\frac{2\pi k}{3}) X(t_0).
\end{aligned}
\end{equation}
In Tsirelson's original formulation, $X$ was the position and $Y$ the momentum of a harmonic oscillator with period $T$. Another natural example is provided by the $J_x$ and $J_y$ components of the angular momentum $\vec{J}$ of a system evolving under the Hamiltonian $H = - (2\pi/T) J_z$. In both these systems, the precession equation above holds at all times; here we relax the requirement and request it to be valid at least at times $\{t_k = kT/3\}_{k\in\mathbbm{Z}}$. This relaxed condition makes it amenable to GPTs, since many such theories with discrete spectra only admit a discrete time evolution \cite{GPT_dynamics}. 

Under this dynamical assumption of precession, the protocol involves many independent rounds. Each round proceeds as follows:

\begin{enumerate}
\item[(1)] Prepare the system in some state.
\item[(2)] Choose at random which duration to wait among the possible $t_k \in \{t_0,t_1,t_2\}$.
\item[(3)] Measure the corresponding $X_k$.
\end{enumerate}
\noindent After many rounds, the score is calculated as
\begin{equation}\label{eq:define-score}
P_3 \coloneqq \frac{1}{3}\sum_{k=0}^2\Big\{
    \underbrace{
        \Pr(X_k>0) +
        \frac{1}{2} \Pr(X_k=0)
    }_{
        \eqqcolon \langle \pos(X_k) \rangle
    }
\Big\}
\end{equation}
Here, $\pos(x)$ is the Heaviside step function. The score $P_3$ is therefore the average likelihood that $X(t)$ was found to be positive at the measured times. The term ${\Pr}(X_k=0)/2$ is due to the convention $\pos(x=0) = 1/2$, which means assigning the score $1/2$ if the system was found exactly at $X=0$.

For classical theory, since the three points $\{X_k\}_{k=0}^2$ in $X$--$Y$ space are equally distributed about a circle of radius $\sqrt{(X_0)^2 + (Y_0)^2}$ centered at the origin, they cannot be all positive or all negative. Therefore, the classical score is upper bounded as $P_3 \leq \mathbf{P}_3^c \coloneqq 2/3$: this is \emph{Tsirelson's original inequality}, following the nomenclature by \citet{plavala2023tsirelson}. \nocite{plavala2023tsirelson} Violation of this inequality was initially shown for the quantum harmonic oscillator (up to $P_3 \gtrsim 0.709$) \cite{OG_tsirelson}, and later for spin angular momentum (up to $P_3 = 3/4$) \cite{precession_protocol}.

In this Letter, we analyze similar bounds for quantum and even more general theories, as will be introduced in the following section.

\section{Observables in General Theories} \label{sec:assumptions}
If they are eventually meant to describe the physical world, general theories should accommodate some concept of periodic oscillation or spatial rotation. Thus, there should also be a notion of precession dynamics in general theories as in Eq.~\eqref{eq:precession-condition}, which is given as a statement about \emph{observables}. What exactly these observables are depends on the theory---for example, they are real-valued random variables in classical theory and self-adjoint operators in quantum theory.

For general theories, we leave open the nature of observables: what mathematical object describes an observable $A$, how measurement outcomes $a \in \mathbb{R}$ are found from $A$, and how outcome probabilities $\Pr(A=a|s)$ are obtained from a description $s$ of the state of the system. Rather, we only require that the expectation value $\ev{A}_s \coloneqq \sum_a a \Pr(A=a|s)$ is a linear function of the observables. That is,
\begin{equation}\label{eq:linearity-of-mean}
\begin{aligned}
\ev{c_1A_1 + c_2A_2}_s = c_1\ev{A_1}_s + c_2\ev{A_2}_s.
\end{aligned}
\end{equation}
It is natural to expect Eq.~\eqref{eq:linearity-of-mean} to hold, insofar as taking linear combinations and expectation values describe our operations rather than properties of nature. Of course, one could always invent a theory where it does not, in which case our results would not be applicable.

That said, for all definitions of observables in finite-dimensional GPTs, which include classical and quantum theories in particular, linearity of the mean is always satisfied \cite{supplementary}. \nocite{GPT_review,hypersphere-qubit,GPT-CV-1,GPT-CV-2,GPT_observables_1,GPT_observables_2,GPT_dynamics,SDP,quartic_QM, Galley2021,McKague} In fact, they may also hold in theories that do not obey the usual restrictions imposed on GPTs: in Supplemental Material, we construct a general theory that does not satisfy the axioms of GPTs, but where the linearity of the mean still holds \cite{supplementary}.

Note that for uniformly precessing observables, Eq.~\eqref{eq:precession-condition} together with Eq.~\eqref{eq:linearity-of-mean} give
\begin{equation}\label{eq:means-precess}
\begin{aligned}
    \ev{X_k}_s &=
        \cos(\frac{2\pi k}{3}) \ev{X_0}_s +
        \sin(\frac{2\pi k}{3}) \ev{Y_0}_s,
\end{aligned}
\end{equation}
which in turn implies that $\sum_{k=0}^2 \ev{X_k}_s = 0$.

\section{Spectrum of Observables}
In this work, we consider systems where the set of outcomes of the measured observable is discrete and finite. For an observable $X_k$, we denote the set of all possible measurement outcomes of $X_k$ as
\begin{equation}\label{eq:spectrum-definition}
\lambda(X_k) = \{-x_{-d_-},\dots,-x_{-2},-x_{-1}, 0, x_{+1},\dots,x_{+d_+}\},
\end{equation}
with $-x_{-d_-} < \dots < -x_{-2} < -x_{-1} < 0 < x_{+1} < \dots < x_{+d_+}$. We will call $\lambda(X) \coloneqq \bigcup_{k=0}^2 \lambda(X_k)$, the set of all possible measurement outcomes of $X$ over all measured times, the \emph{spectrum} of the observable $X$.

Note that the spectrum of a uniformly precessing observable is time independent, since the label of the initial time $t_0$ is an arbitrary choice up to a relabeling. To illustrate this, notice that measuring $X_k$ with the state $s$ produces the same outcomes as measuring $X_j$ with the state $s'$, where the state preparation procedure for $s'$ involves preparing $s$ and waiting the time $(t_k-t_j) \bmod T$. Hence, every measurement outcome of $X_k$ is contained in the spectrum of $X_j$.

More generally, every observable evolving under closed dynamics has a time-independent spectrum in classical, quantum, and generalized probabilistic theories. This is because time evolutions in these theories are described by transformations that preserve the state space, which leaves the measurement outcomes time independent \cite{GPT_dynamics,Muller_Ududec,Galley2021}. Therefore, if $X_k$ is related to $X$ by a time evolution, $\lambda(X) = \lambda(X_k)$ refers to the spectrum of $X$ at any time without any ambiguity.

In addition, note also that the spectrum might not include $0$ (e.g.~in the quantum description of spin-half angular momentum), and $\lambda(X)$ could also be nonnegative or nonpositive. For a quantum observable specified by an operator, the spectrum defined here corresponds to its mathematical spectrum, which is consistent with the usual notion of finite-dimensional quantum systems as discrete variable systems.

An example of a discrete variable system in classical theory is the projection of the hand of an analog clock onto one of its diameters. With the familiar convention of 60 divisions (minutes or seconds) of the clock, the possible measurement outcomes are $\{l\cos(s\pi/30)\}_{s=0}^{30}$, where $l$ is the length of the hand. With the three probing times $t_k = 20k$, Eq.~\eqref{eq:precession-condition} is satisfied for the horizontal diameter $X$ and vertical diameter $Y$.

\section{General Theory-independent Bound}
Let us make the following \emph{physical} assumptions (P):
(P1) The observables $(X,Y)$ are uniformly precessing [Eq.~\eqref{eq:precession-condition}], and (P2) the expectation value is a linear function of the observables [Eq.~\eqref{eq:linearity-of-mean}].

Together, they imply the following \emph{minimal} assumption (M): The sum of the means of $(X_0,X_1,X_2)$ vanishes for all states [$\forall s: \sum_{k=0}^2 \ev{X_k}_s = 0$].

Although (P)$ \implies $(M), we highlight both here because assumptions (P) are physically motivated in oscillating systems, while assumption (M) is minimal for deriving the following theory-independent bound:
\begin{proposition}
In any general theory where assumption (M) is satisfied for the observables $(X_0,X_1,X_2)$, the score $P_3$ as defined in Eq.~\eqref{eq:define-score} is bounded above by
\begin{equation}\label{eq:general-bound}
    P_3 \leq \mathbf{P}_3^G({x_{+}}/{x_{-}}) \coloneqq 
    \begin{cases} 
        \pqty{1+{x_+}/{x_{-}}}^{-1} & 
        \begin{array}{l}
             \text{if $x_{+} < x_{-}$} \\[-0.5ex]
             \text{or $0 \notin \lambda(X)$}
        \end{array} \\[1.5ex]
        1/2 & \text{otherwise,}
    \end{cases}
\end{equation}
where $x_{+} \coloneqq \min\{ x \in \lambda(X) : x > 0\}$, $-x_{-} \coloneqq \min\{ x \in \lambda(X) : x < 0 \}$, and $\lambda(X)$ is the spectrum of $X$.
\end{proposition}
The proof is given in Supplemental Material \cite{supplementary}. The general bound depends solely on the ratio between the minimum positive and negative measurement outcomes, respectively denoted $x_+$ and $-x_-$. While this bound is spectrum dependent, it is nonetheless theory independent, as the spectrum was defined purely operationally as the set of all possible measurement outcomes as in Eq.~\eqref{eq:spectrum-definition}. Notice also that Eq.~\eqref{eq:general-bound} includes the case where the spectrum is nonnegative (nonpositive), for which $x_{-}$ ($x_{+}$) is taken to be undefined, such that ``\emph{otherwise}'' implies $\mathbf{P}^G_3 = 1/2$.

If the general bound is violated, then the assumptions of the precession protocol were wrong. As the condition in Eq.~\eqref{eq:means-precess} that the means precesses uniformly can be operationally verified from the measurement outcomes, the only possibility left is that the seemingly basic assumption about the linearity of the expectation value must be wrong. Since all expectation values are linear for all GPTs, violation of the general bound implies that the given theory does not fall under the framework of GPTs. In other words, what most think to be naturally possible---like always being able to mix states or measurements \cite{GPT_review}---would be falsified.

\section{Saturation by Quantum Theory}
We shall now prove by construction that the general bound is saturated by quantum theory: that is, for a given $x_+$ and $x_-$, it is always possible to construct a uniformly precessing quantum observable that achieves $P_3 = \mathbf{P}_3^G$. However, it was previously known for angular momentum that Tsirelson's inequality can only be violated by spins $j\geq 3/2$ \cite{precession_protocol}, and the same arguments can be extended to show that any quantum system with dimension three or less only obtains the trivial score $P_3=1/2$ for any given pair of precessing variables \cite{supplementary}. Therefore, we consider the following observables of a four-level system
\begin{equation}\label{eq:simplest-quantum-observable}
\begin{aligned}
X &\coloneqq (x_--x_+) \pqty{ \ketbra{1}{2} + \ketbra{2}{1} } \\
    &\qquad{}+{} \sqrt{x_+x_-}\pqty{
    \ketbra{0}{1} + \ketbra{1}{0} + \ketbra{2}{3} + \ketbra{3}{2}
}, \\
Y &\coloneqq -i(x_--x_+)\pqty\big{ \ketbra{1}{2} - \ketbra{2}{1} } \\
    &\qquad{}-{} i\sqrt{x_+x_-}\pqty{
    \ketbra{0}{1} - \ketbra{1}{0} + \ketbra{2}{3} - \ketbra{3}{2}
},
\end{aligned}
\end{equation}
where $x_- > x_+ > 0$. $X$ can be easily diagonalized using standard techniques to find that its spectrum is $\lambda(X) = \{-x_-,-x_+,x_+,x_-\}$ with the eigenstates
\begin{equation}
\begin{aligned}
    \ket{\pm x_-} &\propto \sqrt{x_+}\pqty{\ket{0} \pm \ket{3}} \pm \sqrt{x_-}\pqty{\ket{1} \pm \ket{2}}, \\
    \ket{\pm x_+} &\propto \sqrt{x_-}\pqty{\ket{0} \mp \ket{3}} \pm \sqrt{x_+}\pqty{\ket{1} \mp \ket{2}}.
\end{aligned}
\end{equation}
Under the Hamiltonian $H=\omega\sum_{n=0}^3\ketbra{n}$, the observables $X$ and $Y$ evolves in time as
\begin{equation}
\begin{aligned}
    X(t) &= e^{i H t} X e^{-i H t} =
    \cos(\omega t) X + \sin(\omega t) Y, \\
    Y(t) &= e^{i H t} Y e^{-i H t} =
    \cos(\omega t) Y - \sin(\omega t) X.
\end{aligned}
\end{equation}
By choosing the probing times as $t_k = 2\pi k/3\omega$, Eq.~\eqref{eq:precession-condition} is satisfied, so $X$ and $Y$ indeed precess uniformly.

With these observables, Hamiltonian, and probing times, the state $\ket{\mathbf{P}_3} \coloneqq (\ket{0}-\ket{3})/\sqrt{2}$ achieves the score
\begin{equation}\label{eq:optimal-quantum-score}
\begin{aligned}
P_3
&= \frac{1}{3}\sum_{k=0}^2 \bra{\mathbf{P}_3} \pos(X_k) \ket{\mathbf{P}_3} \\
&= \frac{1}{3}\sum_{k=0}^2 \bra{\mathbf{P}_3} e^{i H t_k} \pqty\big{\ketbra{x_+}{x_+} + \ketbra{x_-}{x_-}} e^{-i H t_k}\ket{\mathbf{P}_3} \\
&= (1+x_+/x_-)^{-1}
= \mathbf{P}_3^G(x_+/x_-),
\end{aligned}
\end{equation}
which saturates the general bound. Notice that $\mathbf{P}_3^G$ can be made arbitrarily close to $1$ by choosing $x_-$ to be arbitrarily large, but one cannot set $x_+=0$, because $x_+$ is by definition the smallest positive value of the spectrum: said differently, setting it to be $0$ amounts to
setting the spectrum to be $\lambda(X) = \{-x_-,0,x_-\}$, for which the bound becomes $\mathbf{P}_3^G = 1/2$.

The best provable upper bound for the maximum violation $\mathbf{P}_3^\infty$ with the quantum harmonic oscillator is $\mathbf{P}_3^\infty \leq 0.730\,822$ \cite{allThreeAnglesTsirelson}. Hence, when $2.72 x_+ \leq x_-$, the state $\ket{\mathbf{P}_3}$ obtains the score $P_3 \gtrsim 0.731 > \mathbf{P}_3^\infty$. That is, discrete quantum systems can provably outperform the harmonic oscillator in the precession protocol.

Of the observables that outperform the harmonic oscillator, a special case is obtained when $x_- = 3x_+$, whereupon $X \propto J_x$ and $Y \propto J_y$ for $j=3/2$. This also shows that the previously found quantum violation of $P_3 = 3/4$ by the angular momentum operators of a spin-$3/2$ particle \cite{precession_protocol} is the maximum possible violation that can be found in any general theory that shares its spectrum.

That said, a direct comparison is difficult between classical and quantum theory for the spin-$3/2$ particle, as there are no uniformly precessing classical observables with the measurement outcomes. We can revisit the analog clock for an example where both the classical and quantum variables share the same spectrum. For some $N$ that is a positive multiple of $6$, consider
\begin{equation}\label{eq:clock-quantum-observable}
\begin{aligned}
    C_x \coloneqq \frac{l}{2}\pqty{S_N^\dag+S_N} \oplus X, 
    \;\,
    C_y \coloneqq \frac{l}{2i}\pqty{S_N^\dag-S_N} \oplus Y;
\end{aligned}
\end{equation}
where $S_N \coloneqq \ketbra{0}{N-1} + \sum_{n=0}^{N-2}\ketbra{n+1}{n}$ is the shift operator, while $X$ and $Y$ are as previously defined with $x_-$ and $x_+$ to be chosen.

The spectrum of the shift operator is well-known \cite{shift-operator}, from which $l(S_N^\dag + S_N)/2$ can be found to have eigenvalues $\{l \cos(2\pi n/N)\}_{n=0}^{N-1}$ with eigenstates
\begin{equation}\label{eq:clock-eigenstates}
    \ket{\psi_n} \propto \sum_{n'=0}^{N-1} e^{i  2\pi nn'/N} \ket{n'}.
\end{equation}
As such, by choosing
\begin{equation}
\begin{aligned}
    x_- &= -\min \{l \cos(2\pi n/N)\}_{n=0}^{N-1} = l, \\
    x_+ &= \min\{l \cos(2\pi n/N) : l \cos(2\pi n/N) > 0\}_{n=0}^{N-1} \\
    &= l\cos[ 2\pi (\lceil N/4 \rceil - 1)/N ],
\end{aligned}
\end{equation}
we have $\lambda(C_x) = \{l \cos(2\pi n/N)\}_{n=0}^{N-1}$, with the extremal values of $\lambda(C_x)$ corresponding exactly to those of $\lambda(X)$. This spectrum is shared by a classical analog clock with $N$ divisions, with $N=60$ being a standard clock face divided into minutes or seconds.

It can also be verified that Eq.~\eqref{eq:precession-condition} is satisfied by $C_x(t_k) \coloneqq U^{\dag k} C_x U^k$ and $C_y(t_k) \coloneqq U^{\dag k} C_y U^k$ where $U \coloneqq \exp(-i \frac{2\pi}{3}\sum_{n}n\ketbra{n})$, so $C_x$ and $C_y$ are uniformly precessing. Meanwhile, the inclusion of $X$ from Eq.~\eqref{eq:simplest-quantum-observable} in Eq.~\eqref{eq:clock-quantum-observable} ensures that $C_x$ achieves the score $\mathbf{P}_3^G$. For the standard clock face, this gives $\mathbf{P}_3^G = (1+\cos(7\pi/15))^{-1} \approx 0.905$. This is hence an example of a quantum observable with a spectrum shared by a classical observable that saturates the general bound.

\section{Probability Spaces of Precessing Variables}
Our findings have some ramifications on the recently introduced study of constrained conditional probability spaces \cite{plavala2023tsirelson}. There, the authors of Ref. \cite{plavala2023tsirelson} considered the set spanned by all possible tuples $(\langle\pos(X_k)\rangle)_{k=0}^2$ achievable in classical theory, which is a strictly smaller set than the full probability space permitted by the algebraic bounds $\forall k : 0 \leq \langle\pos(X_k)\rangle \leq 1$. In the End Matter, we plot the probability spaces occupied by the quantum observables $J_x$ and $C_x$, and compare them to the classical probability subspace. There, we also show that quantum theory can approach the full probability space arbitrarily closely.

\section{Difference between the general theory-independent bound and actual precession}
Notice that the general theory-independent bound is derived only from the form of the observables, without taking into account the requirement that the precession is caused by a time evolution within the theory. Hence, it might seem surprising at first that uniformly precessing systems in quantum theory saturate the upper bound.

Other theories, despite satisfying the linearity of the mean, do not precess in this sense. To illustrate this, let us consider the example of boxworld, also known as generalized nonsignaling theory~\cite{Barrett2007}. If we consider a single system in this theory, a system characterized by $d$ 2-outcome measurements can be represented as a $d$-dimensional hypercube. Reversible transformations on this system are then a subgroup of the symmetry group of the state space and thus in this example discrete~\cite{Muller_Ududec,GPT_dynamics}. This means that we cannot define precession with a continuous parameter $t$ in this case.

One could instead aim at constructing precessing systems with discrete rotations. While there are rotations of order $3$ in a (hyper-)cube, this can not be straightforwardly generalized to boxworld systems characterized by measurements with more than two outcomes. This can be seen most easily as the number of extremal states of a single system is $( \# \text{ outcomes})^{\# \text{ inputs}}$, which generally does not coincide with the number of extremal vertices of a hypercube. Because of this, we were not even able to build interesting precessing systems in this theory.

On the other hand, there are also theories that have been  built as adaptations of quantum theory. These include quantum theory over real Hilbert spaces and Grassmannian systems~\cite{quartic_QM, Galley2021}, which also saturate our bound \cite{supplementary}. We leave for future works to assess whether all systems that accommodate uniform precession and saturate our theory-independent bound are as closely related to quantum theory as these. In this sense, although it fulfills the necessary condition that there is no gap between the quantum and general bounds, this work falls short of providing the first example of a \emph{self-test of quantum theory}~\cite{self-testing1, self-testing2}: the latter remains a conjecture.

\section{Conclusion}
In this Letter, a general theory-independent bound was derived for the precession protocol applied to discrete variable systems. We showed that this bound depends on the spectrum of the measured observables, and that no general theory with finite measurement outcomes can outperform quantum mechanics. A violation of the classical inequality can only be due to incompatible measurements: thus, this protocol defines a figure of merit for which the incompatibility allowed by quantum theory is maximal. Relations to other measures and generalized notions of incompatibility are left for future works.

We have also shown that the previously studied spin-$3/2$ case maximally violates Tsirelson's inequality, and that discrete quantum systems can outperform the quantum harmonic oscillator in the precession protocol. Finally, we related our work to the study of constrained conditional probability spaces, and demonstrated that the quantum set reaches arbitrarily closely to the full unconstrained probability space.

Our work shows that the precession protocol can be used to distinguish quantum theory from any alternate theories that do not also saturate the general theory-independent bound. However, because of the sparse examples of dynamics in general theories, explicit examples of a theory that violate Tsirelson's inequality but do not saturate the bound have eluded us so far. Furthermore, in parallel to the research programs finding underlying physical axioms that recover quantum theory \cite{information-causality}, the lack of a gap between the quantum and general bounds means that the precession condition can be singled out as a possible underlying axiom that recovers certain boundaries of quantum dynamics.

Related work includes the study of \emph{spin-bounded rotation boxes} \cite{rotation-boxes-1,rotation-boxes-2}, where the authors considered general systems transforming under an $\operatorname{SO}(2)$ rotation. Our fundamental assumptions differ, as their assumption is on the \emph{measurement probabilities} (that they are given by a Fourier series with a bounded number of terms), while ours is on the \emph{observables} (that they precess uniformly). A future avenue of research could be to borrow ideas from their approach. Strictly speaking, we have only imposed a $\mathbb{Z}_3$ symmetry on the $X$: what if we imposed an $\operatorname{SO}(2)$ symmetry in the same vein as the authors, or perhaps even $\operatorname{SO}(3)$? The latter restriction would impose that the measured observables $X$ and $Y$ should be components of a three-dimensional vector $(X,Y,Z)$. The general bound we derived here is not saturated by any angular momentum of spin $j>3/2$; nor did we find any other example of vectors in quantum theory that saturated the general bound. It remains to be seen if these imposed symmetries would tighten the bound for vectorial quantities.

It would be interesting to extend this result to the continuous variable case. However, the study of continuous-variable generalized theories~\cite{GPT-CV-1,GPT-CV-2} is still in its infancy, even for GPTs: we were unable to tackle this problem with the existing tools, and thus leave this for the future when the appropriate tools have been developed.

We have also only focused on the Tsirelson-type inequality that was introduced in the original paper, which involves probing the system at three different times. Another possible extension of this work would be to derive similar theory-independent bounds for the Tsirelson-type inequalities that were introduced later, where the system is probed at $K>3$ different times for $K$ odd.

While we have only focused on the precession protocol and assumptions (P) due to the ubiquity of oscillations in physical systems, the same general bound holds in any other scenario, in which the minimal assumption (M) could be motivated. The set of observables $\{X_k\}_{k=0}^2$ that satisfy $\forall s : \sum_{k=0}^2 \ev{X_k}_s = 0$ may have different spectra. The definition and study of such scenarios, as well as the possibility of tightening the bound if the spectra are not the same, are left as open questions.

\section*{Acknowledgments}
This work is supported by the National Research Foundation, Singapore, and A*STAR under its CQT Bridging Grant; by the same National Research Foundation, Singapore, through the National Quantum Office, hosted in A*STAR, under its Centre for Quantum Technologies Funding Initiative (S24Q2d0009); and by the Swiss National Science Foundation (Ambizione PZ00P2\textunderscore208779). We are also grateful to an anonymous referee for pointing out the minimal assumption required for the general bound.

\bibliography{refs}

%apsrev4-2.bst 2019-01-14 (MD) hand-edited version of apsrev4-1.bst
%Control: key (0)
%Control: author (8) initials jnrlst
%Control: editor formatted (1) identically to author
%Control: production of article title (0) allowed
%Control: page (0) single
%Control: year (1) truncated
%Control: production of eprint (0) enabled
\begin{thebibliography}{35}%
\makeatletter
\providecommand \@ifxundefined [1]{%
 \@ifx{#1\undefined}
}%
\providecommand \@ifnum [1]{%
 \ifnum #1\expandafter \@firstoftwo
 \else \expandafter \@secondoftwo
 \fi
}%
\providecommand \@ifx [1]{%
 \ifx #1\expandafter \@firstoftwo
 \else \expandafter \@secondoftwo
 \fi
}%
\providecommand \natexlab [1]{#1}%
\providecommand \enquote  [1]{``#1''}%
\providecommand \bibnamefont  [1]{#1}%
\providecommand \bibfnamefont [1]{#1}%
\providecommand \citenamefont [1]{#1}%
\providecommand \href@noop [0]{\@secondoftwo}%
\providecommand \href [0]{\begingroup \@sanitize@url \@href}%
\providecommand \@href[1]{\@@startlink{#1}\@@href}%
\providecommand \@@href[1]{\endgroup#1\@@endlink}%
\providecommand \@sanitize@url [0]{\catcode `\\12\catcode `\$12\catcode `\&12\catcode `\#12\catcode `\^12\catcode `\_12\catcode `\%12\relax}%
\providecommand \@@startlink[1]{}%
\providecommand \@@endlink[0]{}%
\providecommand \url  [0]{\begingroup\@sanitize@url \@url }%
\providecommand \@url [1]{\endgroup\@href {#1}{\urlprefix }}%
\providecommand \urlprefix  [0]{URL }%
\providecommand \Eprint [0]{\href }%
\providecommand \doibase [0]{https://doi.org/}%
\providecommand \selectlanguage [0]{\@gobble}%
\providecommand \bibinfo  [0]{\@secondoftwo}%
\providecommand \bibfield  [0]{\@secondoftwo}%
\providecommand \translation [1]{[#1]}%
\providecommand \BibitemOpen [0]{}%
\providecommand \bibitemStop [0]{}%
\providecommand \bibitemNoStop [0]{.\EOS\space}%
\providecommand \EOS [0]{\spacefactor3000\relax}%
\providecommand \BibitemShut  [1]{\csname bibitem#1\endcsname}%
\let\auto@bib@innerbib\@empty
%</preamble>
\bibitem [{\citenamefont {Tsirelson}(2006)}]{OG_tsirelson}%
  \BibitemOpen
  \bibfield  {author} {\bibinfo {author} {\bibfnamefont {B.}~\bibnamefont {Tsirelson}},\ }\href@noop {} {\bibinfo {title} {How often is the coordinate of a harmonic oscillator positive?}} (\bibinfo {year} {2006}),\ \Eprint {https://arxiv.org/abs/quant-ph/0611147} {arXiv:quant-ph/0611147 [quant-ph]} \BibitemShut {NoStop}%
\bibitem [{\citenamefont {Zaw}\ \emph {et~al.}(2022)\citenamefont {Zaw}, \citenamefont {Aw}, \citenamefont {Lasmar},\ and\ \citenamefont {Scarani}}]{precession_protocol}%
  \BibitemOpen
  \bibfield  {author} {\bibinfo {author} {\bibfnamefont {L.~H.}\ \bibnamefont {Zaw}}, \bibinfo {author} {\bibfnamefont {C.~C.}\ \bibnamefont {Aw}}, \bibinfo {author} {\bibfnamefont {Z.}~\bibnamefont {Lasmar}},\ and\ \bibinfo {author} {\bibfnamefont {V.}~\bibnamefont {Scarani}},\ }\bibfield  {title} {\bibinfo {title} {Detecting quantumness in uniform precessions},\ }\href {https://doi.org/10.1103/PhysRevA.106.032222} {\bibfield  {journal} {\bibinfo  {journal} {Phys. Rev. A}\ }\textbf {\bibinfo {volume} {106}},\ \bibinfo {pages} {032222} (\bibinfo {year} {2022})}\BibitemShut {NoStop}%
\bibitem [{\citenamefont {Zaw}\ and\ \citenamefont {Scarani}(2023)}]{precession_protocol_anharmonic}%
  \BibitemOpen
  \bibfield  {author} {\bibinfo {author} {\bibfnamefont {L.~H.}\ \bibnamefont {Zaw}}\ and\ \bibinfo {author} {\bibfnamefont {V.}~\bibnamefont {Scarani}},\ }\bibfield  {title} {\bibinfo {title} {Dynamics-based quantumness certification of continuous variables using time-independent hamiltonians with one degree of freedom},\ }\href {https://doi.org/10.1103/PhysRevA.108.022211} {\bibfield  {journal} {\bibinfo  {journal} {Phys. Rev. A}\ }\textbf {\bibinfo {volume} {108}},\ \bibinfo {pages} {022211} (\bibinfo {year} {2023})}\BibitemShut {NoStop}%
\bibitem [{\citenamefont {Jayachandran}\ \emph {et~al.}(2023)\citenamefont {Jayachandran}, \citenamefont {Zaw},\ and\ \citenamefont {Scarani}}]{precession_protocol_CV_ent}%
  \BibitemOpen
  \bibfield  {author} {\bibinfo {author} {\bibfnamefont {P.}~\bibnamefont {Jayachandran}}, \bibinfo {author} {\bibfnamefont {L.~H.}\ \bibnamefont {Zaw}},\ and\ \bibinfo {author} {\bibfnamefont {V.}~\bibnamefont {Scarani}},\ }\bibfield  {title} {\bibinfo {title} {{Dynamics-Based Entanglement Witnesses for Non-Gaussian States of Harmonic Oscillators}},\ }\href {https://doi.org/10.1103/PhysRevLett.130.160201} {\bibfield  {journal} {\bibinfo  {journal} {Phys. Rev. Lett.}\ }\textbf {\bibinfo {volume} {130}},\ \bibinfo {pages} {160201} (\bibinfo {year} {2023})}\BibitemShut {NoStop}%
\bibitem [{\citenamefont {Huynh-Vu}\ \emph {et~al.}(2024)\citenamefont {Huynh-Vu}, \citenamefont {Zaw},\ and\ \citenamefont {Scarani}}]{precession_protocol_DV_ent}%
  \BibitemOpen
  \bibfield  {author} {\bibinfo {author} {\bibfnamefont {K.-N.}\ \bibnamefont {Huynh-Vu}}, \bibinfo {author} {\bibfnamefont {L.~H.}\ \bibnamefont {Zaw}},\ and\ \bibinfo {author} {\bibfnamefont {V.}~\bibnamefont {Scarani}},\ }\bibfield  {title} {\bibinfo {title} {Certification of genuine multipartite entanglement in spin ensembles with measurements of total angular momentum},\ }\href {https://doi.org/10.1103/PhysRevA.109.042402} {\bibfield  {journal} {\bibinfo  {journal} {Phys. Rev. A}\ }\textbf {\bibinfo {volume} {109}},\ \bibinfo {pages} {042402} (\bibinfo {year} {2024})}\BibitemShut {NoStop}%
\bibitem [{\citenamefont {Cirel'son}(1980)}]{Tsirelson-bound}%
  \BibitemOpen
  \bibfield  {author} {\bibinfo {author} {\bibfnamefont {B.~S.}\ \bibnamefont {Cirel'son}},\ }\bibfield  {title} {\bibinfo {title} {Quantum generalizations of bell's inequality},\ }\href {https://doi.org/10.1007/BF00417500} {\bibfield  {journal} {\bibinfo  {journal} {Lett. Math. Phys.}\ }\textbf {\bibinfo {volume} {4}},\ \bibinfo {pages} {93} (\bibinfo {year} {1980})}\BibitemShut {NoStop}%
\bibitem [{\citenamefont {Tsirelson}(1993)}]{PR_tsirelson}%
  \BibitemOpen
  \bibfield  {author} {\bibinfo {author} {\bibfnamefont {B.}~\bibnamefont {Tsirelson}},\ }\bibfield  {title} {\bibinfo {title} {Some results and problems on quantum {B}ell-type inequalities},\ }\href@noop {} {\bibfield  {journal} {\bibinfo  {journal} {Hadronic J. Suppl.}\ }\textbf {\bibinfo {volume} {8}},\ \bibinfo {pages} {329} (\bibinfo {year} {1993})}\BibitemShut {NoStop}%
\bibitem [{\citenamefont {Popescu}\ and\ \citenamefont {Rohrlich}(1994)}]{PR_paper}%
  \BibitemOpen
  \bibfield  {author} {\bibinfo {author} {\bibfnamefont {S.}~\bibnamefont {Popescu}}\ and\ \bibinfo {author} {\bibfnamefont {D.}~\bibnamefont {Rohrlich}},\ }\bibfield  {title} {\bibinfo {title} {Quantum nonlocality as an axiom},\ }\href {https://doi.org/10.1007/BF02058098} {\bibfield  {journal} {\bibinfo  {journal} {Found. Phys.}\ }\textbf {\bibinfo {volume} {24}},\ \bibinfo {pages} {379} (\bibinfo {year} {1994})}\BibitemShut {NoStop}%
\bibitem [{\citenamefont {Lee}\ and\ \citenamefont {Barrett}(2015)}]{Lee_2015}%
  \BibitemOpen
  \bibfield  {author} {\bibinfo {author} {\bibfnamefont {C.~M.}\ \bibnamefont {Lee}}\ and\ \bibinfo {author} {\bibfnamefont {J.}~\bibnamefont {Barrett}},\ }\bibfield  {title} {\bibinfo {title} {Computation in generalised probabilisitic theories},\ }\href {https://doi.org/10.1088/1367-2630/17/8/083001} {\bibfield  {journal} {\bibinfo  {journal} {New J. Phys.}\ }\textbf {\bibinfo {volume} {17}},\ \bibinfo {pages} {083001} (\bibinfo {year} {2015})}\BibitemShut {NoStop}%
\bibitem [{\citenamefont {Weilenmann}\ and\ \citenamefont {Colbeck}(2020{\natexlab{a}})}]{self-testing1}%
  \BibitemOpen
  \bibfield  {author} {\bibinfo {author} {\bibfnamefont {M.}~\bibnamefont {Weilenmann}}\ and\ \bibinfo {author} {\bibfnamefont {R.}~\bibnamefont {Colbeck}},\ }\bibfield  {title} {\bibinfo {title} {{Self-Testing of Physical Theories, or, Is Quantum Theory Optimal with Respect to Some Information-Processing Task?}},\ }\href {https://doi.org/10.1103/PhysRevLett.125.060406} {\bibfield  {journal} {\bibinfo  {journal} {Phys. Rev. Lett.}\ }\textbf {\bibinfo {volume} {125}},\ \bibinfo {pages} {060406} (\bibinfo {year} {2020}{\natexlab{a}})}\BibitemShut {NoStop}%
\bibitem [{\citenamefont {Weilenmann}\ and\ \citenamefont {Colbeck}(2020{\natexlab{b}})}]{self-testing2}%
  \BibitemOpen
  \bibfield  {author} {\bibinfo {author} {\bibfnamefont {M.}~\bibnamefont {Weilenmann}}\ and\ \bibinfo {author} {\bibfnamefont {R.}~\bibnamefont {Colbeck}},\ }\bibfield  {title} {\bibinfo {title} {Toward correlation self-testing of quantum theory in the adaptive {Clauser-Horne-Shimony-Holt} game},\ }\href {https://doi.org/10.1103/PhysRevA.102.022203} {\bibfield  {journal} {\bibinfo  {journal} {Phys. Rev. A}\ }\textbf {\bibinfo {volume} {102}},\ \bibinfo {pages} {022203} (\bibinfo {year} {2020}{\natexlab{b}})}\BibitemShut {NoStop}%
\bibitem [{\citenamefont {Hardy}(2001)}]{Hardy2001}%
  \BibitemOpen
  \bibfield  {author} {\bibinfo {author} {\bibfnamefont {L.}~\bibnamefont {Hardy}},\ }\href@noop {} {\bibinfo {title} {{Quantum Theory From Five Reasonable Axioms}}} (\bibinfo {year} {2001}),\ \Eprint {https://arxiv.org/abs/quant-ph/0101012} {arXiv:quant-ph/0101012 [quant-ph]} \BibitemShut {NoStop}%
\bibitem [{\citenamefont {Barrett}(2007)}]{Barrett2007}%
  \BibitemOpen
  \bibfield  {author} {\bibinfo {author} {\bibfnamefont {J.}~\bibnamefont {Barrett}},\ }\bibfield  {title} {\bibinfo {title} {Information processing in generalized probabilistic theories},\ }\href {https://doi.org/10.1103/PhysRevA.75.032304} {\bibfield  {journal} {\bibinfo  {journal} {Phys. Rev. A}\ }\textbf {\bibinfo {volume} {75}},\ \bibinfo {pages} {032304} (\bibinfo {year} {2007})}\BibitemShut {NoStop}%
\bibitem [{\citenamefont {Plávala}(2022)}]{incompatibility-1}%
  \BibitemOpen
  \bibfield  {author} {\bibinfo {author} {\bibfnamefont {M.}~\bibnamefont {Plávala}},\ }\bibfield  {title} {\bibinfo {title} {Incompatibility in restricted operational theories: connecting contextuality and steering},\ }\href {https://doi.org/10.1088/1751-8121/ac5afe} {\bibfield  {journal} {\bibinfo  {journal} {J. Phys. A}\ }\textbf {\bibinfo {volume} {55}},\ \bibinfo {pages} {174001} (\bibinfo {year} {2022})}\BibitemShut {NoStop}%
\bibitem [{\citenamefont {D’Ariano}\ \emph {et~al.}(2022)\citenamefont {D’Ariano}, \citenamefont {Perinotti},\ and\ \citenamefont {Tosini}}]{incompatibility-2}%
  \BibitemOpen
  \bibfield  {author} {\bibinfo {author} {\bibfnamefont {G.~M.}\ \bibnamefont {D’Ariano}}, \bibinfo {author} {\bibfnamefont {P.}~\bibnamefont {Perinotti}},\ and\ \bibinfo {author} {\bibfnamefont {A.}~\bibnamefont {Tosini}},\ }\bibfield  {title} {\bibinfo {title} {Incompatibility of observables, channels and instruments in information theories},\ }\href {https://doi.org/10.1088/1751-8121/ac88a7} {\bibfield  {journal} {\bibinfo  {journal} {J. Phys. A}\ }\textbf {\bibinfo {volume} {55}},\ \bibinfo {pages} {394006} (\bibinfo {year} {2022})}\BibitemShut {NoStop}%
\bibitem [{\citenamefont {Erba}\ \emph {et~al.}(2024)\citenamefont {Erba}, \citenamefont {Perinotti}, \citenamefont {Rolino},\ and\ \citenamefont {Tosini}}]{incompatibility-3}%
  \BibitemOpen
  \bibfield  {author} {\bibinfo {author} {\bibfnamefont {M.}~\bibnamefont {Erba}}, \bibinfo {author} {\bibfnamefont {P.}~\bibnamefont {Perinotti}}, \bibinfo {author} {\bibfnamefont {D.}~\bibnamefont {Rolino}},\ and\ \bibinfo {author} {\bibfnamefont {A.}~\bibnamefont {Tosini}},\ }\bibfield  {title} {\bibinfo {title} {Measurement incompatibility is strictly stronger than disturbance},\ }\href {https://doi.org/10.1103/PhysRevA.109.022239} {\bibfield  {journal} {\bibinfo  {journal} {Phys. Rev. A}\ }\textbf {\bibinfo {volume} {109}},\ \bibinfo {pages} {022239} (\bibinfo {year} {2024})}\BibitemShut {NoStop}%
\bibitem [{\citenamefont {Branford}\ \emph {et~al.}(2018)\citenamefont {Branford}, \citenamefont {Dahlsten},\ and\ \citenamefont {Garner}}]{GPT_dynamics}%
  \BibitemOpen
  \bibfield  {author} {\bibinfo {author} {\bibfnamefont {D.}~\bibnamefont {Branford}}, \bibinfo {author} {\bibfnamefont {O.~C.}\ \bibnamefont {Dahlsten}},\ and\ \bibinfo {author} {\bibfnamefont {A.~J.}\ \bibnamefont {Garner}},\ }\bibfield  {title} {\bibinfo {title} {On defining the {H}amiltonian beyond quantum theory},\ }\href {https://doi.org/10.1007/s10701-018-0205-9} {\bibfield  {journal} {\bibinfo  {journal} {Found. Phys.}\ }\textbf {\bibinfo {volume} {48}},\ \bibinfo {pages} {982} (\bibinfo {year} {2018})}\BibitemShut {NoStop}%
\bibitem [{\citenamefont {Pl\'avala}\ \emph {et~al.}(2024)\citenamefont {Pl\'avala}, \citenamefont {Heinosaari}, \citenamefont {Nimmrichter},\ and\ \citenamefont {G\"uhne}}]{plavala2023tsirelson}%
  \BibitemOpen
  \bibfield  {author} {\bibinfo {author} {\bibfnamefont {M.}~\bibnamefont {Pl\'avala}}, \bibinfo {author} {\bibfnamefont {T.}~\bibnamefont {Heinosaari}}, \bibinfo {author} {\bibfnamefont {S.}~\bibnamefont {Nimmrichter}},\ and\ \bibinfo {author} {\bibfnamefont {O.}~\bibnamefont {G\"uhne}},\ }\bibfield  {title} {\bibinfo {title} {Tsirelson inequalities: Detecting cheating and quantumness in a single framework},\ }\href {https://doi.org/10.1103/PhysRevA.109.062216} {\bibfield  {journal} {\bibinfo  {journal} {Phys. Rev. A}\ }\textbf {\bibinfo {volume} {109}},\ \bibinfo {pages} {062216} (\bibinfo {year} {2024})}\BibitemShut {NoStop}%
\bibitem [{sup()}]{supplementary}%
  \BibitemOpen
  \href@noop {} {}\bibinfo {note} {See Supplemental Material below for discussions of GPTs, an example of a nonconvex theory, proof of the theory-independent bound, details on semidefinite programming, and saturation of our bound by quantum-inspired GPTs. It includes Refs.~\cite{GPT_review,hypersphere-qubit,GPT_observables_1,GPT_observables_2,GPT-CV-1,GPT-CV-2,SDP,quartic_QM, Galley2021,McKague}}\BibitemShut {NoStop}%
\bibitem [{\citenamefont {Plávala}(2023)}]{GPT_review}%
  \BibitemOpen
  \bibfield  {author} {\bibinfo {author} {\bibfnamefont {M.}~\bibnamefont {Plávala}},\ }\bibfield  {title} {\bibinfo {title} {{General probabilistic theories: An introduction}},\ }\href {https://doi.org/https://doi.org/10.1016/j.physrep.2023.09.001} {\bibfield  {journal} {\bibinfo  {journal} {Phys. Rep.}\ }\textbf {\bibinfo {volume} {1033}},\ \bibinfo {pages} {1} (\bibinfo {year} {2023})}\BibitemShut {NoStop}%
\bibitem [{\citenamefont {Krumm}\ and\ \citenamefont {M{\"u}ller}(2019)}]{hypersphere-qubit}%
  \BibitemOpen
  \bibfield  {author} {\bibinfo {author} {\bibfnamefont {M.}~\bibnamefont {Krumm}}\ and\ \bibinfo {author} {\bibfnamefont {M.~P.}\ \bibnamefont {M{\"u}ller}},\ }\bibfield  {title} {\bibinfo {title} {Quantum computation is the unique reversible circuit model for which bits are balls},\ }\href {https://doi.org/10.1038/s41534-018-0123-x} {\bibfield  {journal} {\bibinfo  {journal} {npj Quantum Inf.}\ }\textbf {\bibinfo {volume} {5}},\ \bibinfo {pages} {7} (\bibinfo {year} {2019})}\BibitemShut {NoStop}%
\bibitem [{\citenamefont {Pl\'avala}\ and\ \citenamefont {Kleinmann}(2022)}]{GPT-CV-1}%
  \BibitemOpen
  \bibfield  {author} {\bibinfo {author} {\bibfnamefont {M.}~\bibnamefont {Pl\'avala}}\ and\ \bibinfo {author} {\bibfnamefont {M.}~\bibnamefont {Kleinmann}},\ }\bibfield  {title} {\bibinfo {title} {{Operational Theories in Phase Space: Toy Model for the Harmonic Oscillator}},\ }\href {https://doi.org/10.1103/PhysRevLett.128.040405} {\bibfield  {journal} {\bibinfo  {journal} {Phys. Rev. Lett.}\ }\textbf {\bibinfo {volume} {128}},\ \bibinfo {pages} {040405} (\bibinfo {year} {2022})}\BibitemShut {NoStop}%
\bibitem [{\citenamefont {Jiang}\ \emph {et~al.}(2024)\citenamefont {Jiang}, \citenamefont {Terno},\ and\ \citenamefont {Dahlsten}}]{GPT-CV-2}%
  \BibitemOpen
  \bibfield  {author} {\bibinfo {author} {\bibfnamefont {L.}~\bibnamefont {Jiang}}, \bibinfo {author} {\bibfnamefont {D.~R.}\ \bibnamefont {Terno}},\ and\ \bibinfo {author} {\bibfnamefont {O.}~\bibnamefont {Dahlsten}},\ }\bibfield  {title} {\bibinfo {title} {Framework for generalized hamiltonian systems through reasonable postulates},\ }\href {https://doi.org/10.1103/PhysRevA.109.032218} {\bibfield  {journal} {\bibinfo  {journal} {Phys. Rev. A}\ }\textbf {\bibinfo {volume} {109}},\ \bibinfo {pages} {032218} (\bibinfo {year} {2024})}\BibitemShut {NoStop}%
\bibitem [{\citenamefont {Filippov}\ \emph {et~al.}(2017)\citenamefont {Filippov}, \citenamefont {Heinosaari},\ and\ \citenamefont {Lepp\"aj\"arvi}}]{GPT_observables_1}%
  \BibitemOpen
  \bibfield  {author} {\bibinfo {author} {\bibfnamefont {S.~N.}\ \bibnamefont {Filippov}}, \bibinfo {author} {\bibfnamefont {T.}~\bibnamefont {Heinosaari}},\ and\ \bibinfo {author} {\bibfnamefont {L.}~\bibnamefont {Lepp\"aj\"arvi}},\ }\bibfield  {title} {\bibinfo {title} {Necessary condition for incompatibility of observables in general probabilistic theories},\ }\href {https://doi.org/10.1103/PhysRevA.95.032127} {\bibfield  {journal} {\bibinfo  {journal} {Phys. Rev. A}\ }\textbf {\bibinfo {volume} {95}},\ \bibinfo {pages} {032127} (\bibinfo {year} {2017})}\BibitemShut {NoStop}%
\bibitem [{\citenamefont {Filippov}\ \emph {et~al.}(2018)\citenamefont {Filippov}, \citenamefont {Heinosaari},\ and\ \citenamefont {Lepp\"aj\"arvi}}]{GPT_observables_2}%
  \BibitemOpen
  \bibfield  {author} {\bibinfo {author} {\bibfnamefont {S.~N.}\ \bibnamefont {Filippov}}, \bibinfo {author} {\bibfnamefont {T.}~\bibnamefont {Heinosaari}},\ and\ \bibinfo {author} {\bibfnamefont {L.}~\bibnamefont {Lepp\"aj\"arvi}},\ }\bibfield  {title} {\bibinfo {title} {Simulability of observables in general probabilistic theories},\ }\href {https://doi.org/10.1103/PhysRevA.97.062102} {\bibfield  {journal} {\bibinfo  {journal} {Phys. Rev. A}\ }\textbf {\bibinfo {volume} {97}},\ \bibinfo {pages} {062102} (\bibinfo {year} {2018})}\BibitemShut {NoStop}%
\bibitem [{\citenamefont {Vandenberghe}\ and\ \citenamefont {Boyd}(1996)}]{SDP}%
  \BibitemOpen
  \bibfield  {author} {\bibinfo {author} {\bibfnamefont {L.}~\bibnamefont {Vandenberghe}}\ and\ \bibinfo {author} {\bibfnamefont {S.}~\bibnamefont {Boyd}},\ }\bibfield  {title} {\bibinfo {title} {{Semidefinite Programming}},\ }\href {https://doi.org/10.1137/1038003} {\bibfield  {journal} {\bibinfo  {journal} {SIAM Rev.}\ }\textbf {\bibinfo {volume} {38}},\ \bibinfo {pages} {49} (\bibinfo {year} {1996})}\BibitemShut {NoStop}%
\bibitem [{\citenamefont {Życzkowski}(2008)}]{quartic_QM}%
  \BibitemOpen
  \bibfield  {author} {\bibinfo {author} {\bibfnamefont {K.}~\bibnamefont {Życzkowski}},\ }\bibfield  {title} {\bibinfo {title} {Quartic quantum theory: an extension of the standard quantum mechanics},\ }\href {https://doi.org/10.1088/1751-8113/41/35/355302} {\bibfield  {journal} {\bibinfo  {journal} {J. Phys. A}\ }\textbf {\bibinfo {volume} {41}},\ \bibinfo {pages} {355302} (\bibinfo {year} {2008})}\BibitemShut {NoStop}%
\bibitem [{\citenamefont {Galley}\ and\ \citenamefont {Masanes}(2021)}]{Galley2021}%
  \BibitemOpen
  \bibfield  {author} {\bibinfo {author} {\bibfnamefont {T.~D.}\ \bibnamefont {Galley}}\ and\ \bibinfo {author} {\bibfnamefont {L.}~\bibnamefont {Masanes}},\ }\bibfield  {title} {\bibinfo {title} {How dynamics constrains probabilities in general probabilistic theories},\ }\href {https://doi.org/10.22331/q-2021-05-21-457} {\bibfield  {journal} {\bibinfo  {journal} {{Quantum}}\ }\textbf {\bibinfo {volume} {5}},\ \bibinfo {pages} {457} (\bibinfo {year} {2021})}\BibitemShut {NoStop}%
\bibitem [{\citenamefont {McKague}\ \emph {et~al.}(2009)\citenamefont {McKague}, \citenamefont {Mosca},\ and\ \citenamefont {Gisin}}]{McKague}%
  \BibitemOpen
  \bibfield  {author} {\bibinfo {author} {\bibfnamefont {M.}~\bibnamefont {McKague}}, \bibinfo {author} {\bibfnamefont {M.}~\bibnamefont {Mosca}},\ and\ \bibinfo {author} {\bibfnamefont {N.}~\bibnamefont {Gisin}},\ }\bibfield  {title} {\bibinfo {title} {{Simulating Quantum Systems Using Real Hilbert Spaces}},\ }\href {https://doi.org/10.1103/PhysRevLett.102.020505} {\bibfield  {journal} {\bibinfo  {journal} {Phys. Rev. Lett.}\ }\textbf {\bibinfo {volume} {102}},\ \bibinfo {pages} {020505} (\bibinfo {year} {2009})}\BibitemShut {NoStop}%
\bibitem [{\citenamefont {M\"uller}\ and\ \citenamefont {Ududec}(2012)}]{Muller_Ududec}%
  \BibitemOpen
  \bibfield  {author} {\bibinfo {author} {\bibfnamefont {M.~P.}\ \bibnamefont {M\"uller}}\ and\ \bibinfo {author} {\bibfnamefont {C.}~\bibnamefont {Ududec}},\ }\bibfield  {title} {\bibinfo {title} {{Structure of Reversible Computation Determines the Self-Duality of Quantum Theory}},\ }\href {https://doi.org/10.1103/PhysRevLett.108.130401} {\bibfield  {journal} {\bibinfo  {journal} {Phys. Rev. Lett.}\ }\textbf {\bibinfo {volume} {108}},\ \bibinfo {pages} {130401} (\bibinfo {year} {2012})}\BibitemShut {NoStop}%
\bibitem [{\citenamefont {Zaw}\ and\ \citenamefont {Scarani}(2024)}]{allThreeAnglesTsirelson}%
  \BibitemOpen
  \bibfield  {author} {\bibinfo {author} {\bibfnamefont {L.~H.}\ \bibnamefont {Zaw}}\ and\ \bibinfo {author} {\bibfnamefont {V.}~\bibnamefont {Scarani}},\ }\href@noop {} {\bibinfo {title} {All three-angle variants of {Tsirelson's} precession protocol, and improved bounds for wedge integrals of {Wigner} functions}} (\bibinfo {year} {2024}),\ \Eprint {https://arxiv.org/abs/2411.03132} {arXiv:2411.03132 [quant-ph]} \BibitemShut {NoStop}%
\bibitem [{\citenamefont {Vourdas}(2004)}]{shift-operator}%
  \BibitemOpen
  \bibfield  {author} {\bibinfo {author} {\bibfnamefont {A.}~\bibnamefont {Vourdas}},\ }\bibfield  {title} {\bibinfo {title} {{Quantum systems with finite Hilbert space}},\ }\href {https://doi.org/10.1088/0034-4885/67/3/R03} {\bibfield  {journal} {\bibinfo  {journal} {Rep. Prog. Phys.}\ }\textbf {\bibinfo {volume} {67}},\ \bibinfo {pages} {267} (\bibinfo {year} {2004})}\BibitemShut {NoStop}%
\bibitem [{\citenamefont {Paw{\l}owski}\ \emph {et~al.}(2009)\citenamefont {Paw{\l}owski}, \citenamefont {Paterek}, \citenamefont {Kaszlikowski}, \citenamefont {Scarani}, \citenamefont {Winter},\ and\ \citenamefont {{\.{Z}}ukowski}}]{information-causality}%
  \BibitemOpen
  \bibfield  {author} {\bibinfo {author} {\bibfnamefont {M.}~\bibnamefont {Paw{\l}owski}}, \bibinfo {author} {\bibfnamefont {T.}~\bibnamefont {Paterek}}, \bibinfo {author} {\bibfnamefont {D.}~\bibnamefont {Kaszlikowski}}, \bibinfo {author} {\bibfnamefont {V.}~\bibnamefont {Scarani}}, \bibinfo {author} {\bibfnamefont {A.}~\bibnamefont {Winter}},\ and\ \bibinfo {author} {\bibfnamefont {M.}~\bibnamefont {{\.{Z}}ukowski}},\ }\bibfield  {title} {\bibinfo {title} {Information causality as a physical principle},\ }\href {https://doi.org/10.1038/nature08400} {\bibfield  {journal} {\bibinfo  {journal} {Nature}\ }\textbf {\bibinfo {volume} {461}},\ \bibinfo {pages} {1101} (\bibinfo {year} {2009})}\BibitemShut {NoStop}%
\bibitem [{\citenamefont {Jones}\ \emph {et~al.}(2023)\citenamefont {Jones}, \citenamefont {Ludescher}, \citenamefont {Aloy},\ and\ \citenamefont {Mueller}}]{rotation-boxes-1}%
  \BibitemOpen
  \bibfield  {author} {\bibinfo {author} {\bibfnamefont {C.~L.}\ \bibnamefont {Jones}}, \bibinfo {author} {\bibfnamefont {S.~L.}\ \bibnamefont {Ludescher}}, \bibinfo {author} {\bibfnamefont {A.}~\bibnamefont {Aloy}},\ and\ \bibinfo {author} {\bibfnamefont {M.~P.}\ \bibnamefont {Mueller}},\ }\href@noop {} {\bibinfo {title} {Theory-independent randomness generation with spacetime symmetries}} (\bibinfo {year} {2023}),\ \Eprint {https://arxiv.org/abs/2210.14811} {arXiv:2210.14811 [quant-ph]} \BibitemShut {NoStop}%
\bibitem [{\citenamefont {Aloy}\ \emph {et~al.}(2024)\citenamefont {Aloy}, \citenamefont {Galley}, \citenamefont {Jones}, \citenamefont {Ludescher},\ and\ \citenamefont {M{\"u}ller}}]{rotation-boxes-2}%
  \BibitemOpen
  \bibfield  {author} {\bibinfo {author} {\bibfnamefont {A.}~\bibnamefont {Aloy}}, \bibinfo {author} {\bibfnamefont {T.~D.}\ \bibnamefont {Galley}}, \bibinfo {author} {\bibfnamefont {C.~L.}\ \bibnamefont {Jones}}, \bibinfo {author} {\bibfnamefont {S.~L.}\ \bibnamefont {Ludescher}},\ and\ \bibinfo {author} {\bibfnamefont {M.~P.}\ \bibnamefont {M{\"u}ller}},\ }\bibfield  {title} {\bibinfo {title} {Spin-bounded correlations: Rotation boxes within and beyond quantum theory},\ }\href {https://doi.org/10.1007/s00220-024-05123-2} {\bibfield  {journal} {\bibinfo  {journal} {Commun. Math. Phys.}\ }\textbf {\bibinfo {volume} {405}},\ \bibinfo {pages} {292} (\bibinfo {year} {2024})}\BibitemShut {NoStop}%
\end{thebibliography}%

\section*{Appendix: Probability Spaces of Precessing Variables}

\begin{figure}[hb]
    \centering
    \includegraphics[width=\columnwidth]{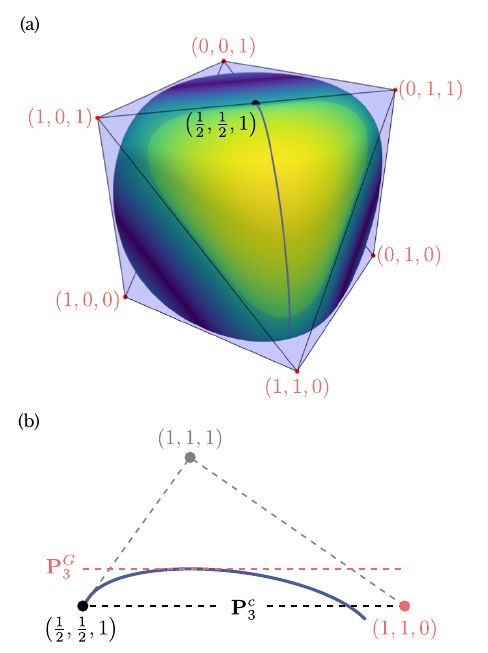}
    \caption{\label{fig:spin32polytope}(a) Probability space $\mathcal{Q}^{(j=3/2)}_{J_x}$ for the outcomes of performing the precession protocol on $J_x$ of a spin-$3/2$ particle. The translucent blue polytope, with its extreme points in red, is the constrained probability space $\mathcal{C}$ that contains the possible tuples $(\langle\pos(X_k)\rangle)_{k=0}^2$ achievable by a classical precessing variable. The original Tsirelson inequality $P_3 = \sum_{k=0}^2\langle\pos(X_k)\rangle/3 \leq 2/3$ is given by the nontrivial facet $\operatorname{Conv}[ \{ (0,1,1), (1,0,1), (1,1,0) \} ]$. The surface is colored according to its distance from the closest nontrivial facet. The classical and quantum sets are both symmetric upon a reflection about the $P_3 = 1/2$ plane. (b) A linecut of $\mathcal{Q}^{(j=3/2)}_{J_x}$ from $(1/2,1/2,1)$ to $(1,1,0)$. The classical and general bounds are the dashed line labeled $\mathbf{P}_3^c$ and $\mathbf{P}_3^G$, respectively.}
\end{figure}

\begin{figure}[hb]
    \centering
    \includegraphics[width=\columnwidth]{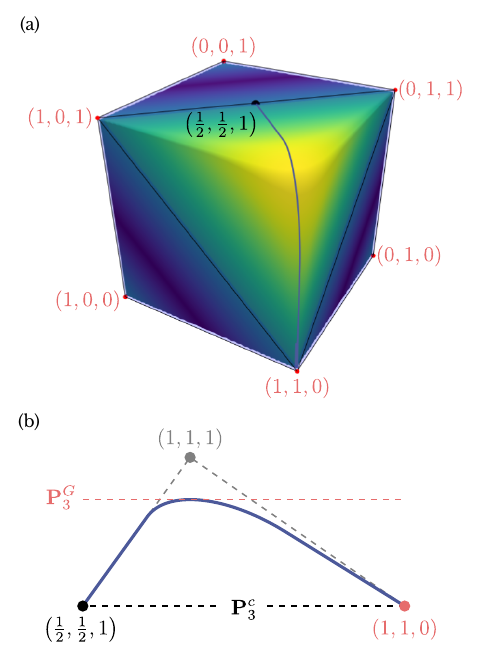}
    \caption{\label{fig:clockSpectrumPolytope}Probability space $\mathcal{Q}_{\text{clock}}^{(N=60)}$ for a quantum observable with the same spectrum as the horizontal and vertical position of the second hand of an analogue clock. The quantum observable saturates the general bound, and the quantum set not only contains the classical set, but also almost entirely fills up the full probability space.}
\end{figure}

In the recently introduced study of constrained conditional probability spaces, \citet{plavala2023tsirelson} considered the probability space spanned by all possible tuples $(\langle\pos(X_k)\rangle)_{k=0}^2$ achievable in classical theory. The full, unrestricted probability space $\mathcal{P}$, which is only constrained by $\forall k : {0 \leq \langle\pos(X_k)\rangle \leq 1}$, is given by the cube
\begin{equation}
    \mathcal{P} \coloneqq \operatorname{Conv}\!\bqty{
        \Bqty{(s_0,s_1,s_2) : s_l \in \{0,1\}}
    },
\end{equation}
where $\operatorname{Conv}(\mathcal{S})$ is the set of all possible convex combinations of elements in $\mathcal{S}$, also known as its convex hull. It was shown that the constraint that $X$ should change signs at least once, which comes from the requirement that it precesses, results in a strictly smaller polytope $\mathcal{C} \subsetneq \mathcal{P}$ for classical observables, where
\begin{equation}
    \mathcal{C} \coloneqq \operatorname{Conv}\!\big[
        \Bqty{ (s_0,s_1,s_2) : s_l \in \{0,1\} }
        \setminus
        \{(0,0,0),(1,1,1)\}
    \big].
\end{equation}
Tsirelson's original inequality then arises from the nontrivial facet
$\operatorname{Conv}[ \{ (0,1,1), (1,0,1), (1,1,0) \} ]$.

The probability space of a given quantum observable $X_k$ can be found using a semidefinite program \cite{supplementary}. The case of $J_x(t_k)$ for the spin-$3/2$ particle is shown in Fig.~\ref{fig:spin32polytope}. While there is no classical analog of an observable with the same spectrum as a spin-$3/2$ particle, we have nonetheless superimposed the classical polytope that must be satisfied by every uniformly precessing classical observable. A substantial volume of the spin-$3/2$ probability space $\mathcal{Q}^{(j=3/2)}_{J_x}$ extrudes out of the nontrivial facet, demonstrating the violation of the classical bound. Meanwhile, the neighborhood of the extremal points of the classical polytope cannot be reached with the quantum system under study.

We can also consider the probability space $\mathcal{Q}_{\text{clock}}^{(N)}$ for the clock observable in Eq.~\eqref{eq:clock-quantum-observable} with $N=60$. For $\ket{\psi_n}$ as defined in Eq.~\eqref{eq:clock-eigenstates}, the tuples $(\bra{\psi_n}\pos[C_x(t_k)]\ket{\psi_n})_{k=0}^2$ form the extreme vertices of $\mathcal{C}$, so their convex combinations replicate the entire classical set. Meanwhile, there are also many points outside the nontrivial facet of $\mathcal{C}$. Hence, $\mathcal{C} \subsetneq \mathcal{Q}_{\text{clock}}^{(N)} \subsetneq \mathcal{P}$, as shown in Fig.~\ref{fig:clockSpectrumPolytope} for $N=60$.

Generally, we have $\overline{\mathcal{Q}}^{(N)}_{\text{clock}} \subseteq \mathcal{Q}^{(N)}_{\text{clock}}$ for any $N$, where
\begin{equation}
\begin{aligned}
    \overline{\mathcal{Q}}^{(N)}_{\text{clock}} \coloneqq \operatorname{Conv}\!\Big[
        \mathcal{C}
        &{}\cup{} \Bqty{
            \pqty{
                \mathbf{P}_3^G,
                \mathbf{P}_3^G,
                \mathbf{P}_3^G
            }
        } \\
        &{}\cup{} \Bqty{
            \pqty{ 
                1-\mathbf{P}_3^G,
                1-\mathbf{P}_3^G,
                1-\mathbf{P}_3^G
            }
        }
    \Big],
\end{aligned}
\end{equation}
since $\ket{\mathbf{P}_3}$ from Eq.~\eqref{eq:optimal-quantum-score} and $e^{-i\pi \sum_{n}\ketbra{n}}\ket{\mathbf{P}_3}$ form the vertices $(\mathbf{P}_3^G)_{k=0}^2$ and $(1-\mathbf{P}_3^G)_{k=0}^2$, respectively.

As $\mathbf{P}_3^G = \{1+{\cos}[ 2\pi (\lceil N/4 \rceil - 1)/N]\}^{-1} \to 1$ when $N \to \infty$, we have $\overline{\mathcal{Q}}^{(N)}_{\text{clock}} \to \mathcal{P}$, which in turn implies that $\overline{\mathcal{Q}}^{(N)}_{\text{clock}} \subseteq \mathcal{Q}^{(N)}_{\text{clock}} \to \mathcal{P}$. This shows that the quantum set can reach arbitrarily close to the full probability space with an observable that has a finite but arbitrarily large number of measurement outcomes.

\clearpage

\phantomsection\addcontentsline{toc}{part}{Supplemental Material for: Tsirelson's Inequality for the Precession Protocol is Maximally Violated by Quantum Theory}

\title{Supplemental Material for: Tsirelson's Inequality for the Precession Protocol is Maximally Violated by Quantum Theory}

\maketitle

\setcounter{section}{0}
\setcounter{figure}{0}
\setcounter{equation}{0}
\renewcommand{\thesection}{S\arabic{section}}
\renewcommand{\thefigure}{S\arabic{figure}}
\renewcommand{\theequation}{S\arabic{equation}}

\section{\label{apd:GPT}Generalized Probabilistic Theories}
Generalized probabilistic theories (GPTs) were introduced to study physical theories operationally, in terms of state preparations, evolutions and measurements~\cite{Hardy2001, Barrett2007}; see~\cite{GPT_review} for a recent review. Both classical and quantum theories are examples of GPTs, although the landscape of GPTs is much richer, including theories like boxworld \cite{Barrett2007} and generalized qubits \cite{hypersphere-qubit}.

In GPTs, possible states live in a state space $\mathcal{S} \subset V$, a convex compact set in a real vector space $V$. The effects of that theory form a convex subset $\mathcal{E} \subseteq \mathcal{E}_{\max} \coloneqq \{ e \in V^* : 0 \leq e(s) \leq 1 \; \forall s \in  \mathcal{S}  \}$. The set $\mathcal{E}$ further has to contain the unique effect $u \in \mathcal{E}$ that satisfies $\forall s \in \mathcal{S}:u(s)=1$ and the zero effect $\forall s \in \mathcal{S}:\mathbf{0}(s) = 0$. Any measurement is made up of a set of effects $\{e^a\}_{a \in \mathbb{A}} \subseteq \mathcal{E}$ such that $\sum_{a \in \mathbb{A}} e^a = u$, where $\mathbb{A} \subset \mathbb{R}$ is the set of measurement outcomes. In this work we only consider measurements with a finite number of outcomes. The probability of obtaining the outcome $a$ given the state $s$ is given by $\Pr(A=a|s) = e^a(s)$.

To our knowledge, there are two notions of observable that have been proposed in the GPT literature. We introduce both notions, and show that both obey the properties that imply our theory-independent bound.

One definition of GPT observables, due to \citet{GPT_dynamics}, specifies them as
\begin{equation}
    A \coloneqq \sum_{a \in \mathbb{A}} a e^a,
\end{equation}
which includes the familiar case of classical observables as random variables and quantum observables as operators. For any state $s \in \mathcal{S}$, the expectation value of $A$ on $s$ is given by $A(s)$, since
\begin{equation}
    A(s) = \sum_{a\in \mathbb{A}} ae^a(s) = \sum_{a\in \mathbb{A}} a \Pr(A=a|s) = \ev{A}_s.
\end{equation}
Observables as defined here are weighted sums of effects, and hence are elements of the vector space $V^*$. As such, expectation values inherit the linearity of the vector space, satisfying $[c_A A + c_B B](s) = c_A A(s) + c_B B(s)$.
% As such, they are equipped with the usual operations of scalar multiplication---satisfying $(cA)(s) = c\cdot A(s)$ for $c \in \mathbb{R}$---and convex combination---satisfying $[\lambda A+ (1-\lambda) B](s) = \lambda A(s) + (1-\lambda) B(s)$ for $0 \leq \lambda \leq 1$.

Meanwhile, \citet{GPT_observables_1,GPT_observables_2} define observables in the sense of a map $A: a \mapsto e^a$, whose expectation value on the state $s$ is similarly given by $\langle A \rangle_s = \sum_{a \in \mathbb{A}} a e^a(s)$. With classical post-processing, each measurement outcome $a$ can be changed independently to $ca$ for some real $c \neq 0$. This leads to a new observable $cA : ca \mapsto e^{ca}$ with $e^{ca} \coloneqq \sum_{a' \in \mathbb{A}} \delta_{a',ca}  e^{a'} = e^{a}$, which has the expectation
\begin{equation}\label{eq:property-scalar}
    \left\langle cA \right\rangle_s
    = \sum_{a \in \mathbb{A}} ca e^a(s)
    = c \left\langle A \right\rangle_s,
\end{equation}
we have the notion of $c A$ as the scalar multiplication of the observable $A$ with $c$.

Furthermore, the convex combination of $A:a \mapsto e^a$ and $B: b \mapsto e^b$ is defined as the mixture $\lambda A + (1-\lambda) B: c \mapsto e^c$, where $\mathbb{C} = \mathbb{A} \cup \mathbb{B}$ and $e^c = \lambda e^a$ if $c \in \mathbb{A}, c \notin \mathbb{B}$, $e^c = (1-\lambda) e^b$ if $c \in \mathbb{B}, c \notin \mathbb{A}$ and $e^c = \lambda e^a + (1-\lambda) e^b$ if $c \in \mathbb{A}, c \in \mathbb{B}$. This is well-defined, as $\lambda e^a$ and $(1-\lambda) e^b$ are all valid effects and $\sum_{a \in \mathbb{A}} \lambda e^a + \sum_{b \in \mathbb{B}} (1-\lambda) e^b = u$, since $\{e^a\}_{a \in \mathbb{A}}$ and $\{e^b\}_{b \in \mathbb{B}}$ each make up a measurement. We then observe directly that 
\begin{equation}\label{eq:property-convex}
\left\langle \lambda  A + (1-\lambda) B \right\rangle_s=\lambda \left\langle A \right\rangle_s + (1-\lambda) \left\langle B \right\rangle_s.
\end{equation}
As such, the convex combinations of observables $\lambda A + (1-\lambda) B$ are themselves observables.

Finally, combining the two properties give
\begin{equation}
\begin{aligned}
\ev{c_1A_1 + c_2A_2}_s
&= \ev{
        \tfrac{|c_1|}{|c_1| + |c_2|} \tilde{A}_1 +
        \tfrac{|c_2|}{|c_1| + |c_2|} \tilde{A}_2
    }_s \\
&\qquad\scriptstyle\text{(where $\tilde{A}_j \coloneqq \operatorname{sgn}(c_j)\pqty{|c_1| + |c_2|} A_j$)}\\
&= \tfrac{|c_1|}{|c_1| + |c_2|} \langle{\tilde{A}_1}\rangle_s +
   \tfrac{|c_2|}{|c_1| + |c_2|} \langle{\tilde{A}_2}\rangle_s \\
&= c_1\ev{A_1}_s + c_2\ev{A_2}_s,
\end{aligned}
\end{equation}
thus this alternate definition of observables also results in expectation values that satisfy linearity.

% In either case, the properties of scalar multiplication and convexity are both satisfied, and together imply the linearity of the expectation value: given observables $A$, $B$ and real scalars $c_A$, $c_B$, we have
% \begin{equation}
%     \ev{c_A A + c_B B}_s = c_A \ev{A}_s + c_B \ev{B}_s.
% \end{equation}
An important consequence
of linearity
for the precession protocol is that Eq.~\eqref{eq:precession-condition} from the main text implies the same equation for the average values:
\begin{equation}
    \ev{X_k} = \cos(\frac{2\pi k}{3}) \ev{X_0} + \sin(\frac{2\pi k}{3}) \ev{Y_0},
\end{equation}
so all GPTs satisfy Eq.~\eqref{eq:means-precess} from the main text.

Notice that GPTs are traditionally formulated only for finite dimensional systems, in the sense that $\operatorname{dim}(V)$ is finite. Recently, there have been proposals also for generalizations to theories with continuous variable observables \cite{GPT-CV-1,GPT-CV-2}, for which toy models have been formulated. We restrict our considerations to the finite dimensional case in this work.

Notice further that in the example of quantum theory, the notion of observables in~\cite{GPT_observables_1,GPT_observables_2} includes implementing positive operator-valued measures, whereas the notion of observables in~\cite{GPT_dynamics} are the operators themselves, where the measurement outcomes are implicitly assumed to be their eigenvalues and effects their eigenprojectors.

\section{\label{apd:GPT-torus}A Nonconvex General Theory}
In this example, we shall construct a general theory that satisfies most of the axioms of GPTs as set out in Sec.~\ref{apd:GPT} except for convexity, while showing that the linearity of the mean value still holds.

For $r_0 < R$, consider the effect space to be an \emph{open} torus in $V^*$
\begin{equation}
\begin{aligned}
\mathcal{E} \coloneqq &\Bigg\{
    \spmqty{
        R\cos\theta + r\cos\theta\cos\phi\\
        R\sin\theta + r\sin\theta\cos\phi\\
        r\sin\phi\\
        \lambda
    }^T :\theta,\phi \in [-\pi,\pi], r \in (0,r_0),\\
    &\quad{}\lambda \in[0,1]
\Bigg\} \cup \Bqty{\spmqty{0\\0\\0\\\lambda}^T :\lambda \in[0,1]},
\end{aligned}
\end{equation}
with the unit and zero effects $u=(0,0,0,1)$ and $\mathbf{0}=(0,0,0,0)$ respectively. The subnormalized state space is defined as $\mathcal{S}_{\leq} \coloneqq \{s\in V: \forall e \in \mathcal{E} : 0 \leq e(s) \leq 1 \}$.

The effect space is clearly nonconvex as certain convex combinations of effects would fall into the ``hole'' of the torus, which, excluding the origin, is not a valid effect. Regardless, this does not impose further restrictions on observables as defined by \citet{GPT_dynamics}, and linearity of the mean follows directly from the properties of the vector space in which the observables live.

On the other hand, for observables as defined by \citet{GPT_observables_1,GPT_observables_2}, the property for scalar multiplication still holds: if $A:a\mapsto e^a$ is a valid observable with $e^a \in \mathcal{E}$, then $cA:ca \mapsto e^{ca}$ with $e^{ca} = e^{a} \in \mathcal{E}$ is still a valid observable for $c \neq 0$, while $cA: a \mapsto \delta_{a,0}u$ for $c=0$.

Meanwhile, consider $A: a \mapsto e^a$ and $B : b \mapsto e^b$ for which $\forall a: e^a \in \mathcal{E}$ and $\forall b: e^b \in \mathcal{E}$ but $\exists a,b: (e^a + e^b)/2 \notin \mathcal{E}$. It might not appear as if $(A+B)/2$ is a valid observable because of the latter fact. However, if we use the scalar multiplication property,
\begin{equation}
    C \coloneqq \frac{1}{2}(A + B)
    = \lambda \underbrace{\frac{A}{2\lambda}}_{\eqqcolon\tilde{A}} +
    (1-\lambda) \underbrace{\frac{B}{2(1-\lambda)}}_{\eqqcolon\tilde{B}},
\end{equation}
we can define the observable $\tilde{A} : \tilde{a} \to e^{\tilde{a}}$ where $\tilde{a} = a/2\lambda$ and $e^{\tilde{a}} = e^a$, and similarly for $\tilde{B}$. For $0<\lambda<1$ large enough, $\lambda e^{\tilde{a}} + (1-\lambda) e^{\tilde{b}}$ will be close enough to $e^a$ so that $\forall \tilde{a},\tilde{b}: \lambda e^{\tilde{a}} + (1-\lambda) e^{\tilde{b}} \in \mathcal{E}$, such that we can now define $C$ in the usual way using $\tilde{A}$ and $\tilde{B}$.

\section{Proof of the Theory-independent Bound}
For a given initial state of the system $s$, we shall denote the probability of measuring $X_k$ and observing $\pm x_{\pm j} \in \lambda(X)$ as $p_{\pm j}(t_k) \coloneqq \Pr(X_k = \pm x_{\pm j}|s)$, henceforth dropping any reference to $s$. With this notation, Eq.~\eqref{eq:define-score} from the main text reads
\begin{equation}
    P_3 = \frac{1}{3}\sum_{k=0}^2\sum_{j=1}^{d_+} p_{+j}(t_k) + \frac{1}{6}\sum_{k=0}^2 p_{0}(t_k).
\end{equation}
Recall that assumption (M) is %From $\sum_{k=0}^2 e^{i2\pi k/3} = 0$, we also have
\begin{equation}\label{eq:means-precess-sum}
\begin{aligned}
    \sum_{k=0}^{2} \ev{X_k} 
    % &= \sum_{k=0}^{2}\bqty{
    %     \cos(\frac{2\pi k}{3}) \ev{X_0} + \sin(\frac{2\pi k}{3}) \ev{Y_0}
    % } \\
    &= 0.
\end{aligned}
\end{equation}
As such, for the special case of nonnegative $\lambda(X)$,
\begin{equation}\label{eq:special-case-nonnegative}
\begin{aligned}
\sum_{k=0}^{2} \ev{X_k} = 0 = \sum_{k=0}^{2} \sum_{j=1}^{d_+} x_{+j} p_{+j}(t_k).
\end{aligned}
\end{equation}
Since $\forall j : x_{+j} > 0$, the only possibility is $\forall j,k:p_{+j}(t_k) = 0$. If $0 \in \lambda(X)$, this implies that $\forall k: p_0(t_k) = 1$ and hence $P_3 = 1/2$. If $0 \notin \lambda(X)$, Eq.~\eqref{eq:means-precess-sum} cannot be satisfied, so
$(X_k)_{k=0}^2$ cannot satisfy assumption (M).
%$X$ is not uniformly-precessing.
An analogous proof shows that $P_3=1/2$ when $\lambda(X)$ is nonpositive and $0\in\lambda(X)$.

We henceforth need only consider the cases where the spectrum contains both positive and negative values, such that $d_\pm \geq 1$. 

\begin{widetext}
Starting from $\ev{X(t)} = \sum_{j=1}^{d_+} x_{+j} p_{+j}(t) - \sum_{j'=1}^{d_-} x_{-j'} p_{-j'}(t)$ and substituting the normalization condition $\sum_{j=1}^{d_+}p_{+j}(t) + \sum_{j'=1}^{d_-}p_{-j'}(t) + p_0(t) = 1$ into $p_{-d_-}$ gives
\begin{equation}
\begin{aligned}
    \ev{X(t)} &= \sum_{j=1}^{d_+} \pqty{x_{+j} + x_{-d_-}} p_{+j}(t) - x_{-d_-} +  x_{-d_-}p_0(t) + \sum_{j'=1}^{d_--1} \pqty{x_{-d_-} - x_{-j'}} p_{-j'}(t).
\end{aligned}
\end{equation}
Since $x_{-d_-} \geq x_{-j'}$ for all $j'$ and $p_{-j'}(t) \geq 0$, we must have $\sum_{j'=1}^{d_--1} (x_{-d_-} - x_{-j'}) p_{-j'}(t) \geq 0$ for the last term, so 
\begin{equation}
\begin{gathered}
    0 \leq \ev{X(t)} - \bqty{ \sum_{j=1}^{d_+} \pqty{x_{+j} + x_{-d_-}} p_{+j}(t) - x_{-d_-} + x_{-d_-}p_0(t) }.
\end{gathered}
\end{equation}
Which can be rearranged as $\sum_{j=1}^{d_+} (x_{+j} + x_{-d_-}) p_{+j}(t) \leq \langle X(t) \rangle + x_{-d_-} - x_{-d_-}p_0(t)$. Therefore,
\begin{equation}
\begin{aligned}
    \ev{\pos[X(t)]} =
    \sum_{j=1}^{d_+}p_{+j}(t) + \frac{1}{2}p_0(t)
    &\leq \sum_{j=1}^{d_+} \frac{x_{+j}+x_{-d_-}}{x_{+1}+x_{-d_-}} p_{+j}(t) + \frac{1}{2}p_0(t) \\
    &\leq \frac{\ev{X(t)} + x_{-d_-} - x_{-d_-}p_0(t)}{x_{+1}+x_{-d_-}} + \frac{1}{2}p_0(t) \\
    &= \frac{\ev{X(t)}}{x_{+1}+x_{-d_-}} + \frac{x_{-d_-}}{x_{+1}+x_{-d_-}} + \frac{x_{+1}-x_{-d_-}}{2(x_{-d_-}+x_{+1})}p_0(t).
\end{aligned}
\end{equation}
If $x_{+1} < x_{-d_-}$, then the last term is negative and can be upper bounded by $0$. Similarly, the last term is zero when $0 \notin \lambda(X)$. Otherwise, $\frac{x_{+1}-x_{-d_-}}{2(x_{-d_-}+x_{+1})}p_0(t) \leq \frac{x_{+1}-x_{-d_-}}{2(x_{-d_-}+x_{+1})}$ from the requirement that $p_0(t) \leq 1$ is a probability. Then,
\begin{equation}
\begin{aligned}
    \ev{\pos[X(t)]} &\leq \frac{\ev{X(t)}}{x_{+1}+x_{-d_-}} +  \begin{cases} 
         \frac{2x_{-d_-}}{2(x_{+1}+x_{-d_-})} & 
        \text{if $x_{+1} < x_{-d_-}$ or $0 \notin \lambda(X)$} \\[1.5ex]
        \frac{2x_{-d_-} + x_{+1} - x_{-d_-}}{2(x_{+1}+x_{-d_-})} & \text{otherwise}
    \end{cases}
\end{aligned}
\end{equation}
Finally, we have
\begin{equation}\label{eq:general-bound-proof}
\begin{aligned}
    P_3 = \frac{1}{3}\sum_{k=0}^{2}\ev{\pos(X_k)} 
    &\leq \frac{1}{3(x_{+1}+x_{-d_-})} \overbrace{\sum_{k=0}^{2}\ev{X_k}}^{0\text{ from Eq.~\eqref{eq:means-precess-sum}}} + \overbrace{\frac{1}{3}\sum_{k=0}^{2}}^{1} \begin{cases} 
         \frac{2x_{-d_-}}{2(x_{+1}+x_{-d_-})} & 
        \text{if $x_{+1} < x_{-d_-}$ or $0 \notin \lambda(X)$} \\[1.5ex]
        \frac{2x_{-d_-} + x_{+1} - x_{-d_-}}{2(x_{+1}+x_{-d_-})} & \text{otherwise}
    \end{cases} \\[1ex]
    &= %\mathbf{P}_3^G(a_{+1}/a_{-d_-}) \coloneqq 
    \begin{cases} 
        (1+x_{+1}/x_{-d_-})^{-1} & 
        \text{if $x_{+1} < x_{-d_-}$ or $0 \notin \lambda(X)$}
        \\
        1/2 & \text{otherwise}
    \end{cases}
    \\[1ex]
    &\eqqcolon \mathbf{P}_3^G(x_{+1}/x_{-d_-}).
\end{aligned}
\end{equation}
Therefore, the general bound depends solely on the ratio between the minimum positive and negative measurement outcomes, respectively $x_{+1} = \min\{ a \in \lambda(X) : a > 0\}$ and $-x_{-d_-} = \min\{ a \in \lambda(X) : a < 0 \}$. Note that Eq.~\eqref{eq:general-bound-proof} nicely includes the special cases: if the spectrum is nonnegative (nonpositive), then $x_{-d_-}$ ($x_{+1}$) is undefined while we must also have $0\in \lambda(X)$ for it to
satisfy assumption (M),
%be uniformly-precessing,
so $x_{+1}/x_{-d_-} \nless 1$ and $\mathbf{P}^G_3 = 1/2$ as previously found.
\end{widetext}

\section{\label{apd:dimension-witness}At Least Four Dimensions are Required for Quantum Systems to Obtain Nontrivial Scores}
Consider two quantum observables $X$ and $Y$ that precess uniformly as
\begin{equation}
\begin{aligned}
    X(t) = \cos(\omega t) X + \sin(\omega t) Y \\
    Y(t) = \cos(\omega t) Y - \sin(\omega t) X
\end{aligned}
\end{equation}
in the Heisenberg picture, with no other parametric dependence on time. Define $2 A(t) \coloneqq X(t) + iY(t)$. Then, the Heisenberg equation reads
\begin{equation}
\comm{A(t)}{H} = i\hbar \frac{d}{dt} A(t) = \hbar\omega A(t).
\end{equation}
If $\ket{E}$ is an eigenstate of $H$ with eigenvalue $E$, then
\begin{equation}
\begin{aligned}
H\pqty{A(t)\ket{E}} &= \pqty{A(t)H-\comm{A(t)}{H}}\ket{E} \\
&= \pqty{E - \hbar\omega}\ket{E} \propto \ket{E-\hbar\omega}.
\end{aligned}
\end{equation}
Clearly, $A(t)$ is an annihilation operator of the Hamiltonian $H$, while $A^\dag(t)$ is a creation operator of $H$. 

If a given observable $O(t) = f(A(t),A^\dag(t))$ is a function of $A(t)$ and $A^\dag(t)$, where $\ket{E}$ and $\ket{E'}$ are eigenstates of $H$ with eigenvalues $E$ and $E'$ respectively, then it can be shown that $\mel{E}{O(t)}{E'} \neq 0 \implies E-E' = \Delta n\hbar\omega$ for some integer $\Delta n$.

To see this, first consider the term $O_{n_1,m_1,\dots,n_N,m_M} = A^{\dag n_1}(t)A^{m_1}(t)A^{\dag n_2}(t)A^{m_2} \dots A^{\dag n_N}(t)A^{m_M}(t)$, where $\{n_j\}_{j=1}^N$ and $\{m_k\}_{k=1}^{M}$ are nonnegative integers. Because of the action of $A(t)$ and $A^\dag(t)$ upon $\ket{E'}$,
\begin{equation}
O_{n_1,m_1,\dots,n_N,m_M}\ket{E'} \propto \lvert{E' + ({\scriptstyle\sum_j}  n_j - {\scriptstyle\sum_k} n_k)\hbar\omega}\rangle.
\end{equation}
As such,
\begin{equation}
\begin{aligned}
&\mel{E}{O_{n_1,m_1,\dots,n_N,m_M}}{E'} \neq 0 \\
&\implies \langle E\lvert{E' + ({\scriptstyle\sum_j} n_j - {\scriptstyle\sum_k} n_k)\hbar\omega}\rangle \neq 0 \\
&\implies E = E' + \pqty\Big{\sum_j n_j - \sum_k n_k}\hbar\omega \\
&\implies E - E' = \pqty\Big{\sum_j n_j - \sum_k n_k}\hbar\omega  \eqqcolon \Delta n\hbar\omega.
\end{aligned}
\end{equation}
By expanding $O(t) = f(A(t),A^\dag(t))$ as
\begin{equation}
O(t) = \hspace{-0.75em}\sum_{n_1,m_1,\dots} c_{n_1,m_1,\dots,n_N,m_M}O_{n_1,m_1,\dots,n_N,m_M}
\end{equation}
for some coefficients $c_{n_1,m_1,\dots,n_N,m_M}$, the proof holds for any function of $A(t)$ and $A^\dag(t)$.

This shows that we can write each state as $\ket{E_n^{j}} = \lvert{E_0^{j} + n\hbar\omega}\rangle$ for some ground state $\lvert{E_0^{j}}\rangle$ and $n\geq0$, where $A(t)\lvert{E_0^{j}}\rangle = 0$, $A^{\dag n}(t)\lvert{E_0^{j}}\rangle \neq 0$, and $A^{\dag n}(t)\lvert{E_0^{j}}\rangle \propto \ket{E_n^{j}}$. With this, a stronger statement is that $\mel{E}{O(t)}{E'}\neq 0 \implies |E\rangle = |{E_n^{j}}\rangle$ and $|E'\rangle = |{E_m^{j}}\rangle$ for some $n,m \geq 0$, which are generated from the same ground state $|{E_0^{j}}\rangle$.

Now, a consequence of this is that every function $O(t)$ of $A(t)$ and $A^\dag(t)$ can always be written in the block diagonal form
\begin{equation}
O(t) = \bigoplus_{j} \Pi_{E_0^{j}}O(t)\Pi_{E_0^{j}},
\end{equation}
where $\Pi_{E_0^{j}}$ projects the observable into the subspace spanned by $\Bqty{A^{\dag n}(t)|E_0^{j}\rangle}_{n=0}^{N}$ where $N$ is the integer such that $A^{\dag}(t)\ket{E_N^j} = 0$.

For the precession protocol, we are interested in the observable $Q_3 = \frac{1}{2}\pqty{\mathbbm{1}+\Delta Q_3}$, where
\begin{equation}
\Delta Q_3 = \frac{1}{3}\sum_{k=0}^{2}\sgn\!\bqty{X\pqty{t=\frac{2\pi k}{3\omega}}}
\end{equation}
with $\sgn(x) \coloneqq 2\Theta(x)-1$. Without loss of generality, let us consider only a single block from the above, as the general case can be recovered by taking a direct sum after analyzing each block. The only possible nonzero matrix elements of $\Delta Q_3$ are
\begin{equation}
\begin{aligned}
\mel{E_n}{\Delta Q_3}{E_m} &= \frac{1}{3} \sum_{k=0}^{2} \mel{E_n}{e^{i \frac{2\pi k H}{3\hbar\omega}}\sgn(X)e^{-i \frac{2\pi k H}{3\hbar\omega}}}{E_m} \\
&= \mel{E_n}{\sgn(X)}{E_m} \sum_{k=0}^{2}\frac{e^{i \frac{2\pi \Delta n k}{3}}}{3} \\
&= \mel{E_n}{\sgn(X)}{E_m}\delta_{0,\Delta n\bmod 3},
\end{aligned}
\end{equation}
where $\Delta n = n-m$, and we have used the sum of the cube roots of unity. In other words,
\begin{equation}
\Delta Q_3 = \bigoplus_{k=0}^{2}\Pi_{k}\sgn(X)\Pi_{k},
\end{equation}
where $\Pi_{k}$ is a projector onto the $\Bqty{\ket{E_{3n+k},\lambda}}_{n \geq 0}$ subspace. Note that we now denote the energy eigenstates as $\ket{E_{3n+k},\lambda}$ where $\lambda$ distinguishes the possibly degenerate states.

As a special case, when there are at most three distinct energy levels $\{E_k\}_{0 \leq k < 3}$, then $\Pi_{k} = \sum_{\lambda} \ketbra{E_k,\lambda}$ projects onto the $k$th eigenenergy subspace. Then,
$\Delta Q_3$ has nonzero elements only for $\mel{E,\lambda}{\Delta Q_3}{E',\lambda'} = \mel{E,\lambda}{\sgn(X)}{E,\lambda'}\delta_{E,E'}$. However,
\begin{equation}
\begin{aligned}
\mel{E,\lambda}{\sgn\!\bqty{X(\tfrac{\pi}{\omega})}}{E,\lambda'} &= \mel{E,\lambda}{\sgn(X)}{E,\lambda'} e^{i\pi (E-E)} \\
= -\mel{E,\lambda}{\sgn(X)}{E,\lambda'} 
&= \mel{E,\lambda}{\sgn(X)}{E,\lambda'}.
\end{aligned}
\end{equation}
The last line implies that $\mel{E,\lambda}{\sgn(X)}{E,\lambda'} = 0$, which means that every matrix element of $\Delta Q_3$ is zero. Therefore, $\Delta Q_3 = 0$ when the energies only take the values  $E_0,E_1,E_{2}$, so $P_3 = 1/2$ for all states. In other words, a nontrivial score of $P_3 > 1/2$ is possible only when there are at least four distinct energy levels, which implies minimally a four-level quantum system.

\section{\label{apd:SDP}Finding the Quantum Probability Space with Semidefinite Programming}
To find the quantum probability space, we are interested in finding all possible tuples $(\langle\pos(X_k)\rangle)_{k=0}^2$ achievable for a given quantum observable $X(t)$ given in the Heisenberg picture. Rewriting this as
\begin{equation}
    (\langle\pos(X_k)\rangle)_{k=0}^2
    = \pqty{\tfrac{1}{2},\tfrac{1}{2},\tfrac{1}{2}} + r \hat{n},
\end{equation}
where $\hat{n} = (
    \sin(\theta)\cos(\phi),
    \sin(\theta)\sin(\phi),
    \cos(\theta)
)$ is a unit vector in spherical coordinates, the problem can now be stated as follows: \emph{For a given $\hat{n}$, maximize $r$}. By finding the maximal $r$ for every direction, the quantum probability space can be plotted.

Writing it in the form
\begin{equation}\label{eq:SDP}
\begin{array}{rl}
    \displaystyle\max_{\rho \succeq 0, r \geq 0} & r \\
    \displaystyle\text{subject to} & \displaystyle\tr[\pos(X_0)\rho] = 1/2 + r\sin(\theta)\cos(\phi) \\
    & \displaystyle\tr[\pos(X_1)\rho] = 1/2 + r\sin(\theta)\sin(\phi) \\
    & \displaystyle\tr[\pos(X_2)\rho] = 1/2 + r\cos(\theta) \\
    & \displaystyle \tr\rho = 1,
\end{array}
\end{equation}
we now recognize this as a semidefinite programming (SDP) problem. SDP problems can be solved numerically using standard techniques, where the global optimum can be obtained to within a desired numerical precision \cite{SDP}.

\section{Quantum-inspired GPTs Can Saturate the Theory-independent Bound} \label{apd:GPT_saturation}
$m \times n$ Grassmannian systems are a family of proposed postquantum theories~\cite{quartic_QM, Galley2021}, where the pure states are of the form $\rho = \mathcal{U} \pqty{\ketbra{\psi}\otimes \mathbbm{1}_n} \mathcal{U}^\dag $, with $\mathcal{U} \in SU(m\times n)$ and $\ket{\psi} \in \mathbb{C}^m$.

However, for every quantum state $\ket{\psi}$ and unitary $U$, the mapping $\ket{\psi} \mapsto \ketbra{\psi}\otimes \mathbbm{1}_n$ and $U \mapsto U \otimes \mathbbm{1}_n$ embeds all possible quantum states and dynamics into the Grassmannian system, with projective measurements also given by the same operator as the states. As such, any quantum expectation will be recovered in Grassmannian systems.

Similarly, any complex quantum systems can be encoded into quantum theory over real Hilbert spaces~\cite{McKague}. Namely, any complex state and unitary can be encoded as follows: $\rho \mapsto \frac{1}{2} ( \rho \otimes \ketbra{i} + \rho^* \otimes \ketbra{-i})$ and $U \mapsto  U \otimes \ketbra{i} + U^* \otimes \ketbra{-i}$, where $^*$ denotes complex conjugation and $\ket{\pm i} \coloneqq ( \ket{0} \pm i\ket{1} )/\sqrt{2}$, which again describes a unitary evolution. The encoding of measurements is analogous to that of unitaries. 
To see also explicitly that this encoding gives the optimal quantum value, notice that in Eq.~\eqref{eq:simplest-quantum-observable} from the main text, the encoded observable is $X^{\rm enc}= X \otimes \mathbbm{1}_2$.

\end{document}